\documentclass{aa}  

\usepackage{graphicx}
\usepackage{txfonts}
\usepackage{natbib}
\usepackage{xcolor}
\usepackage{float}
\bibpunct{(}{)}{;}{a}{}{,}
\newcounter{subtable}

\begin{document} 

\renewcommand{\thetable}{\arabic{table}\alph{subtable}}

\title{Testing abundance-age relations beyond solar analogues with {\it Kepler} LEGACY stars
\thanks{Based on observations made at Institut Pytheas/Observatoire de Haute Provence (CNRS), France. Table \ref{tab_abundance_results} is only available in electronic form at the CDS via anonymous ftp to {\tt cdsarc.u-strasbg.fr (130.79.128.5)} or via {\tt http://cdsweb.u-strasbg.fr/cgi-bin/qcat?J/A+A/???/???}}
}
   \titlerunning{Abundance-age relations in the {\it Kepler} LEGACY sample}
  \authorrunning{T. Morel et al.}
   
     \author{Thierry Morel
         \inst{1}
         \and
         Orlagh L. Creevey
         \inst{2}
	 \and
         Josefina Montalb\'an
         \inst{3}
         \and
         Andrea Miglio
         \inst{3}
	 \and
         Emma Willett
         \inst{3}
         }

  \institute{
  Space sciences, Technologies and Astrophysics Research (STAR) Institute, Universit\'e de Li\`ege, Quartier Agora, All\'ee du 6 Ao\^ut 19c, B\^at. B5C, B4000-Li\`ege, Belgium\\
              \email{tmorel@uliege.be}
              \and
              Universit\'e C\^ote d’Azur, Observatoire de la C\^ote d’Azur, CNRS, Laboratoire Lagrange, Bd de l’Observatoire, CS 34229, 06304 Nice Cedex 4, France
  \and
  School of Physics and Astronomy, University of Birmingham, Edgbaston, Birmingham, B15 2TT, UK
  }

   \date{Received 19 August 2020 ; accepted 6 November 2020}

   \abstract{The prospects of using abundance ratios as stellar age indicators appear promising for solar analogues, but the usefulness of this technique for stars spanning a much wider parameter space remains to be established. We present abundances of 21 elements in a sample of 13 bright FG dwarfs drawn from the {\it Kepler} LEGACY sample to examine the applicability of the abundance-age relations to stars with properties strongly departing from solar. These stars have precise asteroseismic ages that can be compared to the abundance-based estimates. We analyse the well-known binary \object{16 Cyg AB} for validation purposes and confirm the existence of a slight metal enhancement ($\sim$0.02 dex) in the primary, which might arise from planetary formation/ingestion. We draw attention to systematic errors in some widely-used catalogues of non-seismic parameters that may significantly bias asteroseismic inferences. In particular, we find evidence that the ASPCAP $T_\mathrm{eff}$ scale used for the APOKASC catalogue is too cool for dwarfs and that the [Fe/H] values are underestimated by $\sim$0.1 dex. In addition, a new seismic analysis of the early F-type star \object{KIC 9965715} based on our spectroscopic constraints shows that the star is more massive and younger than previously thought. We compare seismic ages to those inferred from empirical abundance-age relations based on ages from PARSEC isochrones and abundances obtained in the framework of the HARPS-GTO program. These calibrations take into account a dependency with the stellar effective temperature, metallicity, and/or mass. We find that the seismic and abundance-based ages differ on average by 1.5-2 Gyrs, while taking into account a dependency with one or two stellar parameters in the calibrations leads to a global improvement of up to $\sim$0.5 Gyr. However, even in that case we find that seismic ages are systematically larger by $\sim$0.7 Gyr. We argue that it may be ascribed to a variety of causes including the presence of small zero-point offsets between our abundances and those used to construct the calibrations or to the choice of the set of theoretical isochrones. The conclusions above are supported by the analysis of literature data for a larger number of {\it Kepler} targets. For this extended sample, we find that incorporating a $T_\mathrm{eff}$ dependency largely corrects for the fact that the abundance-based ages are lower/larger with respect to the seismic estimates for the cooler/hotter stars. Although investigating age dating methods relying on abundance data is worth pursuing, we conclude that further work is needed to improve both their precision and accuracy for stars that are not solar analogues.}


\maketitle
%

\section{Introduction}\label{sect_introduction}
  
Stellar ages play a central role in various fields of astrophysics. Yet, it is one of the most difficult properties to determine with good accuracy in field stars. Although traditionally used in the past to tag young or old stellar populations, [Fe/H] is now recognised not to be a reliable age proxy for stars in the Galactic discs \citep[e.g.][]{edvardsson93}. However, several studies of solar twins/analogues spanning a narrow range of parameters have recently unveiled remarkable correlations between stellar ages (derived from isochrone fitting) and other abundance ratios \citep[e.g.][]{da_silva12,bedell18,nissen15,nissen16,nissen17,nissen20,adibekyan16,spina16,spina18,tucci_maia16,jofre20,lin20}. Abundance ratios of elements produced through different nucleosynthesis channels (e.g. [Y/Mg]) are particularly sensitive to age because the relative amounts of ejecta released in the interstellar medium (ISM) strongly varied along the evolution of the Galaxy \citep[e.g.][]{spina16}. The proposed relations between the abundance ratios and age generally extend over $\sim$10 Gyrs and are relatively tight (scatter of $\sim$1 Gyr or even less for a given abundance ratio) for thin-disc stars.

Although this novel dating technique is potentially powerful, doubts remain as to whether it can be applied to different stellar populations and help tackling some important issues, such as the formation history of our Galaxy. This question is particularly relevant and timely in view of the flood of stellar abundances that are already delivered by several medium- to high-resolution spectroscopic surveys \citep[e.g.][]{buder20,gilmore12}. A wealth of information into the processes that shaped our Galaxy is encoded in these data, but reconstructing the timeline of events is essential. The sensitivity of the abundance-age relations against the stellar properties remains largely unexplored. As a matter of fact, recent studies have warned that care should be exercised when applying the abundance-age relations derived from solar analogues to stars with characteristics (especially metallicity) significantly different from those of the Sun, not belonging to the Galactic thin disc, or even located outside the solar circle \citep[e.g.][]{feltzing17,titarenko19,casali20,delgado_mena19}. The last point is a concern because old stars are believed to have radially migrated over large distances across the Galaxy \citep[e.g.][]{sellwood02}. Although these dependencies cannot be ignored, they are currently neither well understood nor well quantified. Some abundance ratios appear to be relatively insensitive to some of these issues (e.g. [Y/Mg] compared to [Sr/Mg]; \citealt{nissen20}), but the dependency may be quantitatively different depending on the nucleosynthesis details of the pair of chemical elements involved.

Another line of progress is related to the fact that the calibrations are based on isochrone ages that fully rely on the predictions of evolutionary models and are of limited applicability for unevolved dwarfs. Even in favourable cases (e.g. subgiants), the accuracy in the age determination of field stars is severely limited by systematic uncertainties  \citep[see discussion in, e.g.,][]{sahlholdt19}. In contrast, ages from asteroseismology can be derived with good confidence even for stars on or slightly off the zero-age main sequence (ZAMS), and are believed to be precise at the $\sim$10-15\% level \citep[e.g.][]{chaplin13,lebreton14}. In addition, isochrone ages may be affected by chromospheric activity for stars typically younger than 4-5 Gyrs \citep{spina20}. Calibrating the abundance ratios against seismic ages would thus constitute a step forward, but the current samples of seismic targets with very precise abundances are still too limited in size. First attempts in that direction were made by \citet[][hereafter N17]{nissen17} who determined the abundances of 12 elements in 10 solar-metallicity stars with an age of at most 7 Gyrs observed by the {\it Kepler} satellite. Their results support the abundance-isochrone age relations for solar analogues/twins, although a larger scatter in the calibrations is noticeable. One of the reasons may be that their targets are significantly hotter and more evolved than the Sun.

We present in this paper the abundances of 21 elements in a sample of 13 bright {\it Kepler} targets spanning a wide parameter range and with precise seismic ages extending up to $\sim$12 Gyrs to further examine the performance of the abundance-age relations for stars with properties departing from solar. Our abundances probe the main nucleosynthesis production channels (iron-peak, $\alpha$- and neutron-capture) and also include lithium. These prime {\it Kepler} targets are amongst the main-sequence, solar-like stars with the best set of fundamental parameters (e.g. mass or age) known from asteroseismology. Our spectroscopic constraints complement the seismic data and will also aid further theoretical modelling. 

\section{Selection of targets and basic properties}\label{sect_targets}

\begin{table*}[h!]
  \caption{Basic properties of the targets. The determination of the stellar parameters is described in Sects.~\ref{sect_seismic_data} and \ref{sect_atmospheric_parameters}.}
\label{tab_properties} 
\scriptsize
\centering
\begin{tabular}{ll|lccl|cccr}
\hline\hline                                          
KIC ID                & Other IDs                                               & \multicolumn{1}{l}{Spectral type} & \multicolumn{1}{c}{$V$}  & \multicolumn{1}{c}{Binary status} & \multicolumn{1}{l}{Population} & \multicolumn{1}{|c}{$T_\mathrm{eff}$ [K]} & \multicolumn{1}{c}{$\log g$ [cgs]} & \multicolumn{1}{c}{$\xi$ [km s$^{-1}$]} & \multicolumn{1}{c}{$[$Fe/H$]$}\\
\hline
\object{KIC 3656476}  &                                                         & G5 IV   & 9.55                 & ...                      & thin disc (?)  & 5680$\pm$15  & 4.233$\pm$0.003 & 1.38$\pm$0.05 &   0.27$\pm$0.03\\
\object{KIC 5184732}  & \object{HD 182756}                                      & G1 V    & 8.31                 & ...                      & thin disc      & 5855$\pm$24  & 4.266$\pm$0.006 & 1.36$\pm$0.05 &   0.41$\pm$0.04\\
\object{KIC 6106415}  & \object{HD 177153}                                      & G0      & 7.21                 & ...                      & thin disc      & 6035$\pm$19  & 4.296$\pm$0.002 & 1.51$\pm$0.08 & --0.05$\pm$0.03\\
\object{KIC 6603624}  &                                                         & G8 IV-V & 9.19                 & ...                      & thin disc (?)  & 5625$\pm$20  & 4.325$\pm$0.004 & 1.26$\pm$0.07 &   0.27$\pm$0.04\\
\object{KIC 7871531}  &                                                         & G5 V    & 9.44                 & SB1/SB2\tablefootmark{a} & thin disc      & 5510$\pm$15  & 4.477$\pm$0.005 & 1.08$\pm$0.05 & --0.19$\pm$0.03\\
\object{KIC 7970740}  & \object{HD 186306}                                      & G9 V    & 7.96                 & ...                      & thick disc (?) & 5365$\pm$15  & 4.545$\pm$0.003 & 1.04$\pm$0.06 & --0.47$\pm$0.03\\
\object{KIC 8006161}  & \object{HD 173701}                                      & G8 V    & 7.52                 & ...                      & thin disc      & 5415$\pm$23  & 4.497$\pm$0.002 & 1.15$\pm$0.13 &   0.35$\pm$0.08\\
\object{KIC 8694723}  &                                                         & G0 IV   & 8.92                 & ...                      & thin disc      & 6295$\pm$56  & 4.112$\pm$0.004 & 1.37$\pm$0.27 & --0.41$\pm$0.06\\
\object{KIC 8760414}  &                                                         & G0 IV   & 9.3\tablefootmark{b} & ...                      & thick disc     & 5985$\pm$35  & 4.329$\pm$0.003 & 1.48$\pm$0.43 & --0.95$\pm$0.05\\ 
\object{KIC 9965715}  &                                                         & F2 V    & 9.21                 & SB2\tablefootmark{a}     & thin disc      & 6335$\pm$40  & 4.280$\pm$0.004 & 1.83$\pm$0.20 & --0.29$\pm$0.04\\ 
\object{KIC 12069424} & \object{16 Cyg A}, \object{HR 7503}, \object{HD 186408} & G1.5 V  & 5.95                 & SB1                      & thin disc      & 5800$\pm$13  & 4.293$\pm$0.002 & 1.34$\pm$0.03 &   0.11$\pm$0.02\\
\object{KIC 12069449} & \object{16 Cyg B}, \object{HR 7504}, \object{HD 186427} & G3 V    & 6.20                 & SB1                      & thin disc      & 5750$\pm$15  & 4.358$\pm$0.004 & 1.30$\pm$0.03 &   0.08$\pm$0.02\\
\object{KIC 12317678} & \object{HD 234998}                                      & F5      & 8.75                 & SB1/SB2\tablefootmark{a} & thin disc      & 6550$\pm$112 & 4.061$\pm$0.011 & 1.13$\pm$0.50 & --0.19$\pm$0.12\\
\hline
\end{tabular}
\tablefoot{
  \tablefoottext{a}{The SB2 classification is based on the analysis of near-IR APOGEE spectra by \citet{el_badry18b}, but note that we do not detect the signature of the secondaries in our optical spectra.}
  \tablefoottext{b}{In $R$ band.}}
\end{table*}

Our targets are drawn from the so-called {\it Kepler} LEGACY sample made up of a total of 66 solar-like dwarfs and subgiants \citep{lund17,silva_aguirre17}. They have all been the subject of intensive and long-duration (up to 4 years) observations by the {\it Kepler} satellite, and are currently the stars with the best seismic parameters available. In particular, as discussed in Sect.~\ref{sect_seismic_data}, two independent studies derived the ages to 5-10\% precision through a thorough modelling of these data. Because the amplitude of solar-like oscillations is proportional to stellar luminosity, this sample is biased towards stars hotter and more evolved than the Sun. However, stars at the end of the core-hydrogen burning phase that exhibit modes of mixed character were excluded \citep[][]{lund17,silva_aguirre17}. 

 To adequately investigate the dependency of the abundance-age relations as a function of [Fe/H] \citep[e.g.][]{feltzing17,skuladottir19,casali20}, we first choose stars in \citet{buchhave15} with a metallicity deviating by more than 2$\sigma$ (i.e. 0.2 dex) from solar. Our stars span a wide range of [Fe/H] values: from --0.95 to +0.41. Given the limited aperture size of the telescope, the final selection of the targets was mainly driven by signal-to-noise (S/N) considerations, and 13 stars with $V$ $<$ 10 mag could eventually be observed. Three near-solar metallicity stars (\object{KIC 6106415} and \object{16 Cyg AB}) with high-precision abundance studies in the literature (N17, and references therein) were included in this sample for validation purposes. Except \object{16 Cyg B} \citep{cochran97}, none of our targets is known to host planets or is a {\it Kepler} Object of Interest (KOI). 

 Very precise radial velocities (RVs) are obtained from our instrument cross-correlation (CCF) data (a G2 mask was used). A comparison with the line-of-sight (LOS) values of \citet{lund17} reveals clear RV changes in \object{KIC 7871531} and \object{KIC 12317678} for which we obtain values larger by $\sim$0.7 and 20.4 km s$^{-1}$, respectively.\footnote{The LOS values of \citet{lund17} for these two stars are based on spectroscopic data obtained in early June 2014: see {\it Kepler} Community Follow-up Observing Program (CFOP) website at \url{https://cfop.ipac.caltech.edu/home/}.} The variations for the former are small, but significant at the $\sim$6$\sigma$ level. We therefore identify these two stars as single-lined (SB1) binaries. The binarity of our targets has recently been investigated on the basis of Apache Point Observatory Galactic Evolution Experiment (APOGEE) data. First, \citet{price_whelan18} applied on the DR14 dataset a method optimised for sparse multi-epoch RV observations, but classified \object{KIC 7871531} as single. On the other hand, \citet{el_badry18b} used another approach and fitted DR13 spectra to single out RV variables and stars with composite spectra (SB2). Three of our targets are flagged as single stars (\object{KIC 3656476}, \object{KIC 6603624}, and \object{KIC 8694723}), whereas three others are classified as SB2's (\object{KIC 7871531}, \object{KIC 9965715}, and \object{KIC 12317678}). No significant RV variations are apparent in our CCF data of \object{KIC 9965715}, which were acquired over two consecutive nights. The analysis of \citet{el_badry18b} confirms the binary nature of \object{KIC 7871531} and \object{KIC 12317678}, although we do not detect the signature of the secondaries in our optical spectra. The APOGEE observations cover the near-infrared (near-IR) $H$-band, and are therefore sensitive to cool companions. We computed the absolute magnitudes of the primaries in the $V$ band, $M_\mathrm{V}$, assuming {\it Gaia} DR2 parallaxes \citep[][]{gaia18}. Although the $E$($B-V$) values are compatible with zero within the uncertainties, we also  used reddening estimates from the 3D maps of \citet{lallement18}\footnote{See online tool at \url{https://stilism.obspm.fr/}.}. For the secondaries, we used the solar-metallicity isochrones of \citet{spada13} with a mixing-length parameter, $\alpha$ = 1.875, to estimate $M_\mathrm{V}$ assuming the seismic masses and ages of the primaries discussed in Sect.~\ref{sect_seismic_data}, along with the mass ratios of \citet{el_badry18b}. We conclude that the companions do not contribute to more than $\sim$10\% to the flux in the optical. No visual companions have been detected in ten of our targets (including \object{KIC 7871531} and \object{KIC 12317678}) from adaptive optics imaging in the visible \citep{schonhut_stasik17}. If confirmed, the composite nature of the spectra is expected to bias our abundances by about 0.05 dex at most \citep{el_badry18a}. Leaving the visual binary 16 Cyg aside, there are to our knowledge no signs of binarity in the other targets. For \object{KIC 8006161}, in particular, no RV changes are detected in our data secured two days apart. This lack of variability is supported by the long-term CORAVEL monitoring of \citet{halbwachs18}.

 We used {\it Gaia} DR2 astrometric and proper motion data \citep[][]{gaia18}, along with our RVs, to compute the space components of the targets relative to the local standard of rest \citep[see][]{morel03}. Following \citet{bensby14}, we then estimated the probability that the stars belong to a given component of the Galaxy. We find compelling evidence that \object{KIC 8760414} displays kinematic properties consistent with a thick-disc membership (it is about 7 times less likely to belong to the halo). The situation is much more ambiguous based on kinematical information alone for three other stars with an age above 8 Gyrs. We find that they are about twice more likely to belong to the thin disc. However, given that kinematical properties do not allow a clean separation between the populations \citep[e.g.][]{bensby14}, additional criteria --- including the abundance pattern and the age --- must be considered. As \object{KIC 8760414}, \object{KIC 7970740} exhibits enhanced abundances of the $\alpha$ elements\footnote{Defined as the unweighted mean of the Mg, Si, and Ti abundances.} ([$\alpha$/Fe]$\sim$+0.2 dex; see Sect.~\ref{sect_results}) at a level more compatible with the high-$\alpha$ sequence defined by thick-disc stars in the [Fe/H]-[$\alpha$/Fe] plane \citep[e.g.][]{reddy06}. Its chemical properties therefore also support a thick-disc membership. We will consider below that these two stars, which turn out to be the oldest and most metal-poor objects in our sample, are members of the thick disc. It can be noted that their age (10.7 and 12.0 Gyrs; Sect.~\ref{sect_seismic_data}) is consistent with this Galactic component having formed about 11 Gyrs ago \citep{silva_aguirre18,miglio20,montalban20}. The two other thick-disc candidates (\object{KIC 3656476} and \object{KIC 6603624}) have large metallicities, but solar abundances of the $\alpha$ elements. This indicates that they do not belong to the population of metal-rich, $\alpha$-enhanced stars (the so-called h$\alpha$mr stars) claimed to be distinct by \citet{adibekyan11}. It is likely that these metal-rich, solar [$\alpha$/Fe] stars with an age of 8-9 Gyrs migrated from the inner Galaxy \citep[e.g.][]{miglio20}. The rest of the sample are clearly thin-disc members.

 Our spectral analysis might be affected by chromospheric activity, although the impact is, as expected, much larger for (very) young objects \citep[e.g.][]{flores16,galarza19}. As recently shown by \citet{spina20}, the effect on spectral lines can be significant for stars with $\log R_\mathrm{HK}^{\prime}$ $\gtrsim$ --5.0, where $R_\mathrm{HK}^{\prime}$ is an activity proxy determined through measurements of the \ion{Ca}{ii} H+K emission-line fluxes. A homogeneous determination of the activity level is provided for 11 of our targets by \citet{brewer16}. The $\log R_\mathrm{HK}^{\prime}$ values are found to tightly cluster in a range roughly solar (from --5.21 to --4.89), even for the only two targets that are much younger than the Sun (the early-F stars \object{KIC 9965715} and \object{KIC 12317678}; see Sect.~\ref{sect_seismic_data}). A much higher activity level at the time of our observations cannot be ruled out, as illustrated by the strong activity cycle of \object{KIC 8006161} \citep{kiefer17,karoff18},\footnote{Note that our spectroscopic observations of \object{KIC 8006161} were accidentally obtained during an activity minimum \citep{karoff18}.} but activity is not expected to be a major concern in our sample largely dominated by stars with an age similar to that of the Sun or older. 

The basic properties of our targets are summarised in Table \ref{tab_properties}. 
 
\section{Observations}\label{sect_observations}

The observations were secured with the \'echelle, fibre-fed SOPHIE spectrograph installed at the 1.93-m telescope of the Observatoire de Haute Provence (OHP, France) during the period 10-12 July 2018. The observations were carried out in the object+sky configuration. The spectra in High-Efficiency (HE) mode have a resolving power, $R$, of about 40,000 and a complete wavelength coverage from about 3875 to 6945 \AA. Two to three exposures were obtained for the faintest targets and co-added according to their S/N. The final S/N per pixel ranges from 213 to 350 at $\sim$5550 \AA, with a median of 289.

For the solar spectrum to be used as reference, we averaged the six HE observations of asteroids (one of Ceres and five of Vesta) in the SOPHIE archives with a S/N above 100 (the mean S/N is 168) and a satisfactory blaze correction. Observations of a point-like source have to be preferred \citep[e.g.][]{gray00}, while co-adding spectra from various reflecting bodies is not an issue \citep[e.g.][]{bedell14}.

Initial processing steps were carried out with the SOPHIE reduction pipeline (DRS). All spectra were corrected for RV shifts based on the DRS CCF data. The spectrograph is simultaneously fed through two circular optical fibres, one illuminated by the target and the other one by the nearby sky. The spectrum obtained through the latter aperture was used for the sky subtraction of the stellar exposure. Finally, the mean spectra were split up into segments with a typical length of about 70 \AA \ and a wavelength coverage carefully chosen in such a way that the continuum level could be adequately fit with low-order cubic spline or Legendre polynomials. To ensure the highest level of homogeneity, the same wavelength bounds and fitting functions were used for all stars. Standard tasks implemented in the IRAF\footnote{{\tt IRAF} is distributed by the National Optical Astronomy Observatories, operated by the Association of Universities for Research in Astronomy, Inc., under cooperative agreement with the National Science Foundation.} software were used for the final reduction steps. 

 \section{Asteroseismic data}\label{sect_seismic_data}

 We make use below of the age, mass, $M$, and surface gravity, $\log g$, determined from asteroseismology. Two teams independently performed a seismic modelling of the oscillation frequencies and/or their separation ratios estimated from the full-length, short-cadence {\it Kepler} dataset \citep{lund17}. While the analysis of \citet[][hereafter C17]{creevey17} is based on a single pipeline \citep[{\tt AMP}: Asteroseismic Modeling Portal;][]{metcalfe09}, \citet[][hereafter SA17]{silva_aguirre17} followed another approach and used six different modelling procedures. As discussed by SA17, the results provided by the various pipelines are consistent despite the variety of codes employed or the different sensitivity of the seismic diagnostics to surface effects, for instance. We thus assume in the following the unweighted mean of all their individual values. It can be noted that C17 only used frequency ratios that are largely insensitive to the uncertain modelling of the near-surface layers. Stellar parameters of {\it Kepler} dwarfs and subgiants determined through a detailed modelling of the full frequency dataset have to be preferred over those inferred from the global seismic quantities because they are more precise by a factor of a few. Notwithstanding, there is no evidence for the LEGACY stars for systematic differences significantly exceeding the uncertainty level \citep[][]{serenelli17}. 

\begin{figure}[h!]
  \centering
  \includegraphics[trim=90 160 90 125,clip,width=1.0\hsize]{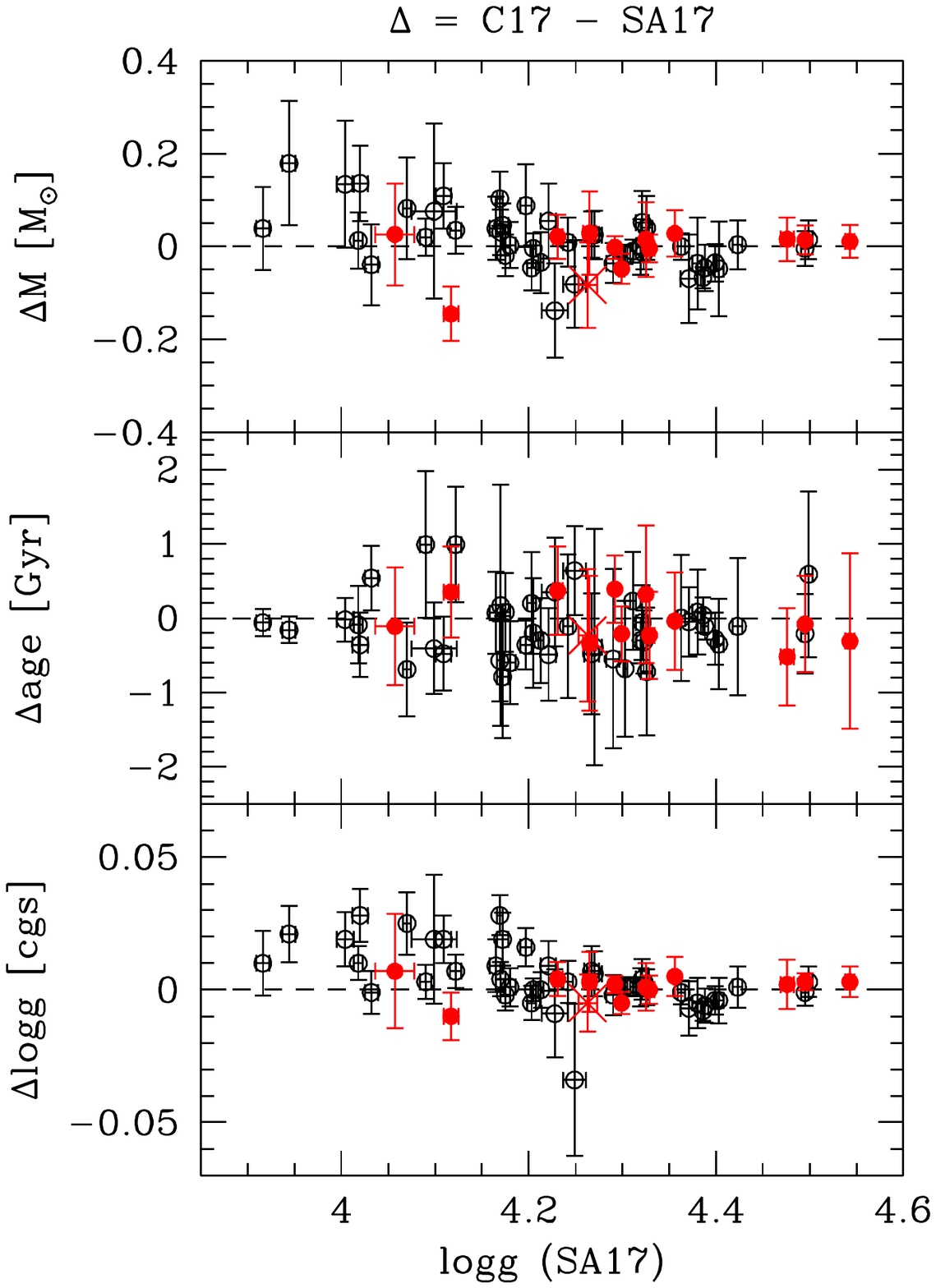}
\caption{Comparison for the {\it Kepler} LEGACY sample between the stellar mass ({\it top panel}), age ({\it middle panel}) and surface gravity ({\it bottom panel}) determined by C17 and SA17. The differences for a given parameter are the values of C17 minus those of SA17. The stars in our sample are shown as red, filled circles. \object{KIC 9965715} is shown for completeness as a cross.}
\label{fig_comparison_seismic_parameters}
\end{figure}

 There is an overall good agreement between the parameters of interest determined by C17 and SA17 (see Fig.~\ref{fig_comparison_seismic_parameters}), but a systematic difference for $M$ and $\log g$ is apparent for the most evolved {\it Kepler} LEGACY stars with $\log g$ $\lesssim$ 4.2. It is beyond the scope of this paper to investigate in depth the origin of these slight discrepancies, which may stem from, e.g., details of the fitting scheme/optimisation procedures or choice of input physics. We simply note that the main differences between the models in SA17 and C17 is the exploration of initial helium abundance and mixing length. Some of the pipelines of SA17 fix the mixing length or restrict it, whereas others impose a chemical enrichment law. Their {\tt C2kSMO} method \citep{lebreton14} is the closest to that employed by C17. An enrichment law is not imposed in C17, and high initial helium mass fractions, $Y_i$, and low masses are occasionally obtained (e.g. \object{KIC 8694723} and \object{KIC 9965715} for which $Y_i$ = 0.309 and 0.310, respectively). As a test, we have fixed $Y_i$ near to 0.257 and 0.267, and the new masses are revised upwards to 1.08 and 1.09 M$_{\sun}$. In any case, this $M$-$Y_i$ correlation does not impact the age. Furthermore, the relatively small differences observed for our sample between the results obtained by C17 and SA17 have no impact on our conclusions. Given that there is also no obvious reason to prefer one set of values over the other, we simply adopt in the following the average values quoted in Table \ref{tab_seismic_data}. 

\begin{table*}[h!]
\caption{Summary of seismic parameters for the sample. The parameters obtained by C17 and SA17 for \object{KIC 9965715} are given for completeness: our updated values can be found in the bottom part of the table.}
\label{tab_seismic_data} 
\scriptsize
\centering
\begin{tabular}{l|ccc|ccc|ccc}
\hline\hline
\multicolumn{5}{c}{} & \multicolumn{1}{c}{{\bf Literature values}} & \multicolumn{4}{c}{} \\
Name                     & \multicolumn{3}{|c|}{$M$ [M$_{\sun}$]} &  \multicolumn{3}{|c|}{Age [Gyr]} & \multicolumn{3}{|c}{$\log g$ [cgs]} \\
                          & C17 & SA17 & Adopted     & \multicolumn{1}{c}{C17} & \multicolumn{1}{c}{SA17} & \multicolumn{1}{c|}{Adopted}     & C17 & SA17 & Adopted \\
\hline
\object{KIC 3656476}                      & 1.101$\pm$0.025 & 1.080$\pm$0.040 & 1.091$\pm$0.024 &  8.88$\pm$0.41 &  8.51$\pm$0.43 &  8.70$\pm$0.30 & 4.235$\pm$0.004 & 4.231$\pm$0.005 & 4.233$\pm$0.003\\
\object{KIC 5184732}                      & 1.247$\pm$0.071 & 1.218$\pm$0.055 & 1.232$\pm$0.045 &  4.32$\pm$0.85 &  4.66$\pm$0.31 &  4.49$\pm$0.45 & 4.268$\pm$0.009 & 4.265$\pm$0.007 & 4.266$\pm$0.006\\
\object{KIC 6106415}                      & 1.039$\pm$0.021 & 1.087$\pm$0.025 & 1.063$\pm$0.016 &  4.55$\pm$0.28 &  4.76$\pm$0.24 &  4.65$\pm$0.18 & 4.294$\pm$0.003 & 4.299$\pm$0.003 & 4.296$\pm$0.002\\
\object{KIC 6603624}                      & 1.058$\pm$0.075 & 1.043$\pm$0.028 & 1.050$\pm$0.040 &  8.66$\pm$0.68 &  8.34$\pm$0.63 &  8.50$\pm$0.46 & 4.326$\pm$0.008 & 4.325$\pm$0.004 & 4.325$\pm$0.004\\
\object{KIC 7871531}                      & 0.834$\pm$0.021 & 0.818$\pm$0.042 & 0.826$\pm$0.023 &  8.84$\pm$0.46 &  9.36$\pm$0.47 &  9.10$\pm$0.33 & 4.478$\pm$0.006 & 4.476$\pm$0.007 & 4.477$\pm$0.005\\
\object{KIC 7970740}                      & 0.768$\pm$0.019 & 0.757$\pm$0.030 & 0.762$\pm$0.018 & 10.53$\pm$0.43 & 10.84$\pm$1.10 & 10.68$\pm$0.59 & 4.546$\pm$0.003 & 4.543$\pm$0.005 & 4.545$\pm$0.003\\
\object{KIC 8006161}                      & 1.000$\pm$0.030 & 0.986$\pm$0.009 & 0.993$\pm$0.016 &  4.57$\pm$0.36 &  4.65$\pm$0.54 &  4.61$\pm$0.32 & 4.498$\pm$0.003 & 4.495$\pm$0.001 & 4.497$\pm$0.002\\
\object{KIC 8694723}                      & 1.004$\pm$0.036 & 1.149$\pm$0.046 & 1.077$\pm$0.029 &  4.85$\pm$0.22 &  4.50$\pm$0.57 &  4.67$\pm$0.31 & 4.107$\pm$0.004 & 4.117$\pm$0.008 & 4.112$\pm$0.004\\
\object{KIC 8760414}                      & 0.814$\pm$0.011 & 0.818$\pm$0.028 & 0.816$\pm$0.015 & 11.88$\pm$0.34 & 12.11$\pm$0.48 & 12.00$\pm$0.29 & 4.329$\pm$0.002 & 4.329$\pm$0.005 & 4.329$\pm$0.003\\
\object{KIC 9965715}                      & 1.005$\pm$0.033 & 1.088$\pm$0.086 & \multicolumn{1}{c|}{...} &  3.29$\pm$0.33 &  3.52$\pm$0.83 &  \multicolumn{1}{c|}{...} & 4.258$\pm$0.004 & 4.263$\pm$0.010 & \multicolumn{1}{c}{...}\\
\object{KIC 12069424}                     & 1.072$\pm$0.013 & 1.074$\pm$0.021 & 1.073$\pm$0.012 &  7.36$\pm$0.31 &  6.97$\pm$0.33 &  7.17$\pm$0.23 & 4.294$\pm$0.001 & 4.292$\pm$0.003 & 4.293$\pm$0.002\\
\object{KIC 12069449}                     & 1.038$\pm$0.047 & 1.010$\pm$0.016 & 1.024$\pm$0.025 &  7.05$\pm$0.63 &  7.09$\pm$0.18 &  7.07$\pm$0.33 & 4.361$\pm$0.007 & 4.356$\pm$0.002 & 4.358$\pm$0.004\\
\object{KIC 12317678}                     & 1.373$\pm$0.030 & 1.347$\pm$0.105 & 1.360$\pm$0.055 &  2.30$\pm$0.20 &  2.41$\pm$0.77 &  2.35$\pm$0.40 & 4.064$\pm$0.005 & 4.057$\pm$0.021 & 4.061$\pm$0.011\\
\multicolumn{1}{c|}{} & \multicolumn{4}{c}{} & \multicolumn{1}{c}{{\bf This study}} & \multicolumn{4}{c}{} \\
Name                  & \multicolumn{3}{|c|}{$M$ [M$_{\sun}$]} & \multicolumn{3}{|c|}{Age [Gyr]} & \multicolumn{3}{|c}{$\log g$ [cgs]} \\
                                        & {\tt ASTEC+ADIPLS}                    & {\tt CLES+LOSC}                 & Adopted                      & \multicolumn{1}{c}{{\tt ASTEC+ADIPLS}} & \multicolumn{1}{c}{{\tt CLES+LOSC}} & \multicolumn{1}{c|}{Adopted} & {\tt ASTEC+ADIPLS}                  & {\tt CLES+LOSC}             & Adopted \\
\hline
\object{KIC 9965715}  & \multicolumn{1}{c}{1.07$\pm$0.04}     & \multicolumn{1}{c}{1.20$^{+0.01}_{-0.13}$}     & \multicolumn{1}{c|}{1.13$\pm$0.05} & \multicolumn{1}{c}{2.8$\pm$0.4}        & \multicolumn{1}{c}{2.41$^{+1.0}_{-0.14}$}         & \multicolumn{1}{c|}{2.6$\pm$0.4} & \multicolumn{1}{c}{4.280$\pm$0.004} & \multicolumn{1}{c}{4.279$^{+0.001}_{-0.01}$} & \multicolumn{1}{c}{4.280$\pm$0.004}\\
\hline
\end{tabular}
\end{table*}

A different procedure was adopted for \object{KIC 9965715} because the significantly different and more reliable spectroscopic constraints we obtain (as argued in Sect.~\ref{sect_validation_parameters_literature}) led us to redetermine the seismic parameters of this star. Two independent modelling approaches referred to in the following as {\tt ASTEC+ADIPLS} and {\tt CLES+LOSC} were performed. Full details are provided in Appendix \ref{sect_seismic_analysis_KIC9965715}. The best-fit parameters obtained are shown in the bottom part of Table \ref{tab_seismic_data}, and the straight mean values adopted in the following. Compared to C17 and SA17, we find a mass revised upwards by $\sim$7\% and that the star is $\sim$30\% younger.

The comparison between the results of various pipelines suggests that the seismic ages used in our study are precise. This is supported by the fact that, for instance, inversion techniques provide an age for 16 Cyg AB \citep[in the range 7.0-7.4 Gyrs;][]{buldgen16} fully compatible with the values adopted. However, we caution that the accuracy of seismic ages cannot be firmly evaluated, although they appear to be relatively robust against the choice of the input physics when detailed seismic information is available \citep[e.g.][]{lebreton14}. It is reassuring that both C17 and SA17 closely reproduced the solar age to much better than $\sim$10\%, but this level of accuracy may not be representative of the whole sample.

\section{Abundance analysis}\label{sect_abundance_analysis}

Except $\log g$ (see above), the atmospheric parameters and abundances of 21 chemical elements were determined from a line-by-line differential analysis relative to the Sun, plane-parallel, 1D MARCS model atmospheres \citep{gustafsson08} and the 2017 version of the line-analysis software {\tt MOOG} originally developed by \citet{sneden73}. 

The line list is taken from \citet{reddy03}, as it was shown that the abundance analysis described below leads to an effective temperature, $T_\mathrm{eff}$, in good agreement for the G2 and K1 main-sequence stars \object{$\alpha$ Cen AB} \citep{morel18} and the G0 subgiant $\beta$ Hyi (unpublished) with interferometric-based estimates \citep{kervella17,heiter15}. In addition, [Fe/H] is within the range of the commonly accepted values for these three stars \citep[e.g.][]{jofre14}. 

The equivalent widths (EWs) were measured manually assuming Gaussian profiles (multiple fits were used for well-resolved blends). Features with an unsatisfactory fit or significantly affected by telluric features based on the atlas of \citet{hinkle00} were discarded. Strong spectral features with RW = $\log$ (EW/$\lambda$) $>$ --4.80 were also excluded provided that a sufficient number of lines are left after this operation. These lines are best avoided for different reasons, e.g., they lie on the non-linear part of curve of growth or are formed in the upper atmosphere and particularly sensitive to chromospheric activity \citep{spina20}.

\subsection{Determination of atmospheric parameters}\label{sect_atmospheric_parameters}

A ``constrained'' analysis whereby the surface gravity is fixed to the very precise seismic value (Table \ref{tab_seismic_data}) was enforced. The model parameters ($T_\mathrm{eff}$, microturbulence, [Fe/H], and mean abundance ratio of the $\alpha$ elements with respect to iron, [$\alpha$/Fe]) were iteratively modified until the following conditions are simultaneously fulfilled: (1) the \ion{Fe}{i} abundances exhibit no trend with RW; (2) the mean abundances derived from the \ion{Fe}{i} and \ion{Fe}{ii} lines are identical; and (3) [Fe/H] and [$\alpha$/Fe] are consistent with the model values. The $\alpha$-element abundance of the model varied depending on [Fe/H] following \citet{gustafsson08}. For instance, [$\alpha$/Fe] = +0.2 for [Fe/H] = --0.4. Although excitation balance of iron is not formally fulfilled in our constrained analysis, it can be noted that the slope between the \ion{Fe}{i} abundances and the lower excitation potential (LEP) deviates from zero by less than $\sim$2$\sigma$ in our sample. The only clear exception is \object{KIC 6603624} for which it appears that the adopted seismic $\log g$ is lower by $\sim$0.3 dex than the value that would be inferred from spectroscopy. For the solar analysis, $T_\mathrm{eff}$ and $\log g$ where held fixed to 5777 K and 4.44, respectively, whereas the microturbulence, $\xi$, was left as a free parameter.

Following N17, we corrected the \ion{Fe}{i} abundances from departures from local thermodynamic equilibrium (LTE). We made use of {\tt Spectrum Tools}\footnote{{\tt http://nlte.mpia.de}.} to interpolate the non-LTE corrections as a function of the stellar parameters. The calculations are described in \citet{bergemann12} and are available for a representative set of lines (about two thirds of the total). This was deemed unnecessary for the \ion{Fe}{ii}-based abundances because non-LTE effects are negligible. Even for \ion{Fe}{i}, the differential corrections are small (less than 0.02 dex), and taking them into account has very little impact on the parameters derived through iron ionisation balance.

Freezing $\log g$ to the seismic value has a much more profound effect on our results. We find that relaxing this constraint would result in $\log g$, $T_\mathrm{eff}$ and [Fe/H] values larger on average by 0.11 dex, 50 K and 0.03 dex, respectively. Our choice of a constrained analysis is motivated by the fact that severe discrepancies between the spectroscopic $\log g$ and the reference value are occasionally observed on a star-to-star basis (e.g., up to 0.5 dex in the extreme case of \object{KIC 12317678}). There is still no consensus among the community as to whether such an analysis has to be preferred for seismic targets \cite[e.g.][]{doyle17}. A constrained analysis is undoubtedly more precise, but whether it is also more accurate is unclear. Our study does not shed new light on this issue. We simply note, as argued in Sect.~\ref{sect_validation_parameters_interferometry}, that a comparison with the handful of long-baseline interferometric measurements available lends some support to the cooler $T_\mathrm{eff}$ scale resulting from the constrained analysis. This is consistent with the recent claim based on a larger sample of interferometric benchmark targets that a constrained analysis is more accurate, especially for F-type stars \citep[][]{gent20}.

\subsection{Determination of chemical abundances}\label{sect_chemical_abundances}

Our line list for elements other than iron is taken from \citet{reddy03}. However, for some key elements (Mg, Al, and Zn) we substituted this line list with that of \citet{melendez14} because not enough lines are included. This was also done for Co because not all lines in \citet{reddy03} have hyperfine structure (HFS) data available. HFS and isotopic splitting were taken into account for Sc, V, Mn, Co, and Cu using atomic data from the Kurucz database\footnote{Available at {\tt http://kurucz.harvard.edu/linelists.html}.} and assuming the Cu isotopic ratio of \citet{rosman98}. The barium abundance is solely based on \ion{Ba}{II} $\lambda$5853 for which HFS and isotopic splitting can be safely neglected \citep[e.g.][]{mashonkina99}. It is supported by our own tests on the Sun using the atomic data of \citet{prochaska00}.

\begin{figure}[h!]
\centering
\includegraphics[trim=40 270 205 180,clip,width=0.7\hsize]{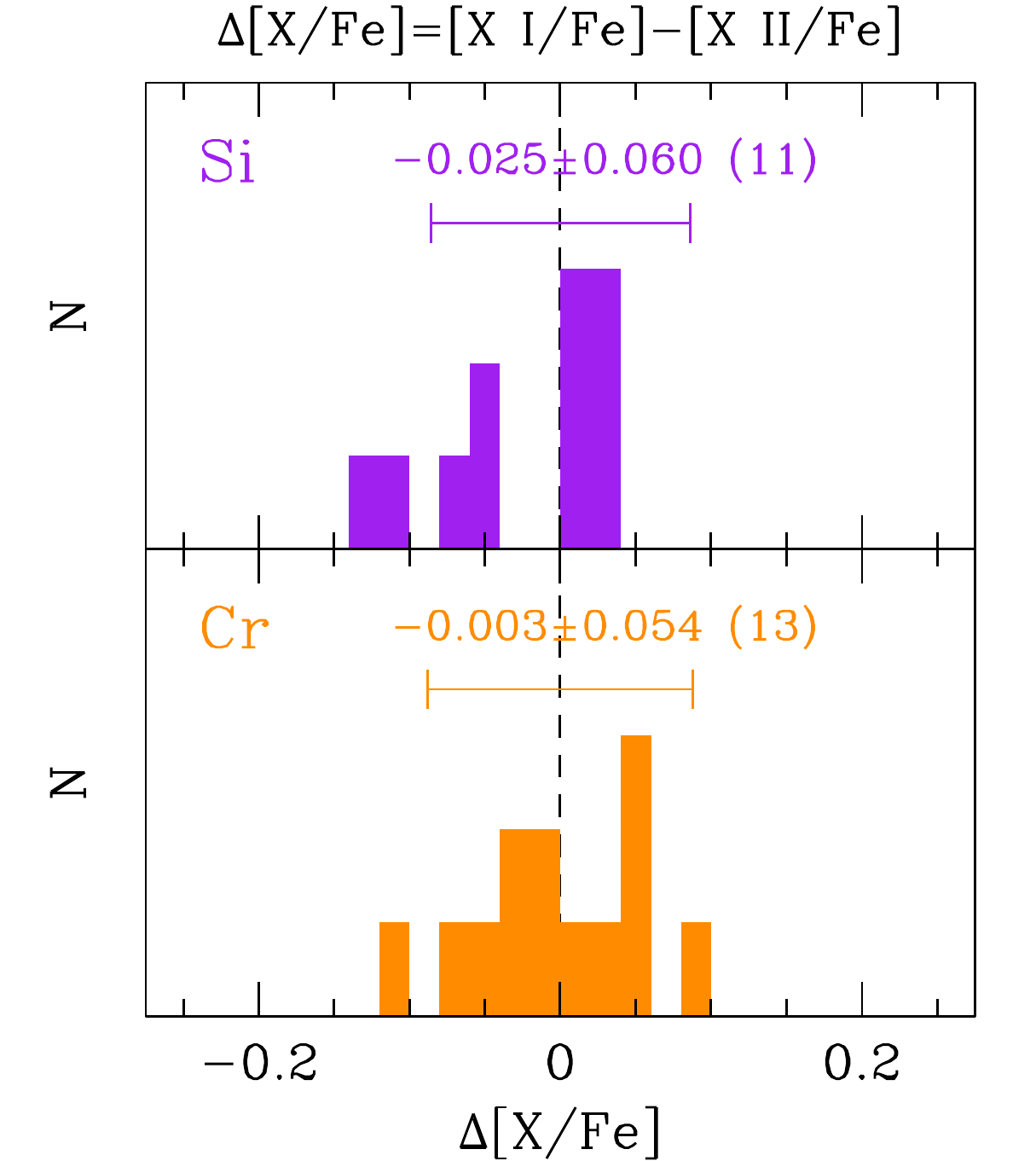}
\caption{Comparison for Si ({\it top panel}) and Cr ({\it bottom panel}) between the mean abundances yielded by the neutral and singly-ionised ion. The mean difference is given in each panel (the number in brackets is the number of stars the calculation is based on). A representative error bar is also shown.}
\label{fig_comparison_abundances_Si_Cr}
\end{figure}

We do not discuss further the abundances derived for \ion{Si}{ii} and \ion{Cr}{ii} because they are based on a single feature, whereas those of the corresponding neutral species rely on up to 6 lines. However, despite the uncertainty plaguing the \ion{Si}{ii}- and \ion{Cr}{ii}-based abundances, ionisation balance is fulfilled for these two elements in the vast majority of cases (Fig.~\ref{fig_comparison_abundances_Si_Cr}).

The determination of the lithium abundance relied on a spectral synthesis of \ion{Li}{i} $\lambda$6708 \citep[for details, see][]{morel14}. This line is confidently detected in 7 targets; upper limits are provided for the others. The rotational and macroturbulent broadening values are taken from \citet{brewer16}. For \object{KIC 8694723} and \object{KIC 8760414} that are not included in this study, we adopted the $v\sin i$ values of \citet{bruntt12} and computed the macroturbulence from the calibrations as a function of $T_\mathrm{eff}$ of \citet{bruntt10}. Tests using the broadening parameters from the works above, but also from \citet{doyle14}, show that the exact choice of these quantities is irrelevant. For the Sun, we obtain an absolute lithium abundance, $\log \epsilon_\sun$(Li)=+1.08, which is in excellent agreement with the recommended photospheric solar value \citep[e.g.][]{asplund09}. For the five stars with a firm detection in common with \citet{bruntt12} and \citet{tucci_maia19}, there is a very good agreement (within 0.1 dex) without any evidence for a systematic offset (see also Sect.~\ref{sect_validation_abundances}). Our non detection for five stars is also consistent with the findings of \citet{bruntt12} and \citet{beck17}.

For a number of reasons, we do not attempt to correct our abundances for the combined effect of departures from LTE and time-dependent convection that would require in principle to drop the rudimentary assumption of 1D. First, performing full 3D, non-LTE radiative transfer calculations is extremely tedious. As a result, theoretical predictions are currently only available for a few elements and very often do not span the whole parameter space of our sample \citep[][]{amarsi19a,amarsi17,nordlander17,gallagher20,bergemann19,mott20}. To our knowledge, spectral line formation under non-LTE has not even been investigated for some key elements in the context of our study (e.g. Y). Correcting for either non-LTE {\it or} 3D effects is not necessarily recommended because the corrections might be of opposite sign and counterbalance each other \citep[e.g.][]{amarsi19b}. As a result, there is no guarantee that the corrected abundances are more accurate. Second, and more importantly, the abundance-age relationships that are the main focus of this paper are built on the assumptions of 1D and LTE \citep[e.g.][]{delgado_mena17}. Should our abundance data be used in a wider context, we caution that care should be exercised for some elements that are known to be particularly sensitive to non-LTE and/or 3D effects, especially in the metal-poor regime \citep[e.g. Mn;][]{bergemann19}.

\section{Results}\label{sect_results}

The atmospheric parameters and chemical abundances are given in Tables \ref{tab_properties} and \ref{tab_abundance_results}, respectively. The random uncertainties are computed following common practice \citep[see][]{morel18}. For the abundances, the line-to-line scatter, $\sigma_\mathrm{int}$, and the uncertainties arising from errors in the stellar parameters are added in quadrature. For iron, $\sigma_\mathrm{int}$ typically amounts to 0.04 dex. For the elements with a single diagnostic line, we assume $\sigma_\mathrm{int}$ = 0.05 dex. The uncertainties are much larger for the three metal-poor, hottest stars because of the paucity and weakness of the diagnostic \ion{Fe}{i} lines.

We explored to what extent the choice of another family of model atmospheres would affect our results. Namely, we repeated the analysis using ATLAS9 Kurucz models \citep{castelli03} for 5 representative stars spanning the parameter space of our sample in terms of $T_\mathrm{eff}$, $\log g$, and [Fe/H]: \object{KIC 7871531}, \object{KIC 8006161}, \object{KIC 8694723}, \object{KIC 8760414}, and \object{16 Cyg A}. We find in all cases very small $T_\mathrm{eff}$ differences not exceeding 5 K. Although the use of Kurucz models leads to systematically lower microturbulent velocities (on average by 0.035 km s$^{-1}$), this is paralleled in the Sun. As a net result, we find that all the abundance ratios differ by a negligible amount (virtually in all cases below 0.01 dex).

\section{Validation of stellar parameters}\label{sect_validation_parameters}

\subsection{Comparison with other spectroscopic studies}\label{sect_validation_parameters_literature}

\begin{figure}[h!]
\centering
\includegraphics[trim=125 145 130 140,clip,width=0.85\hsize]{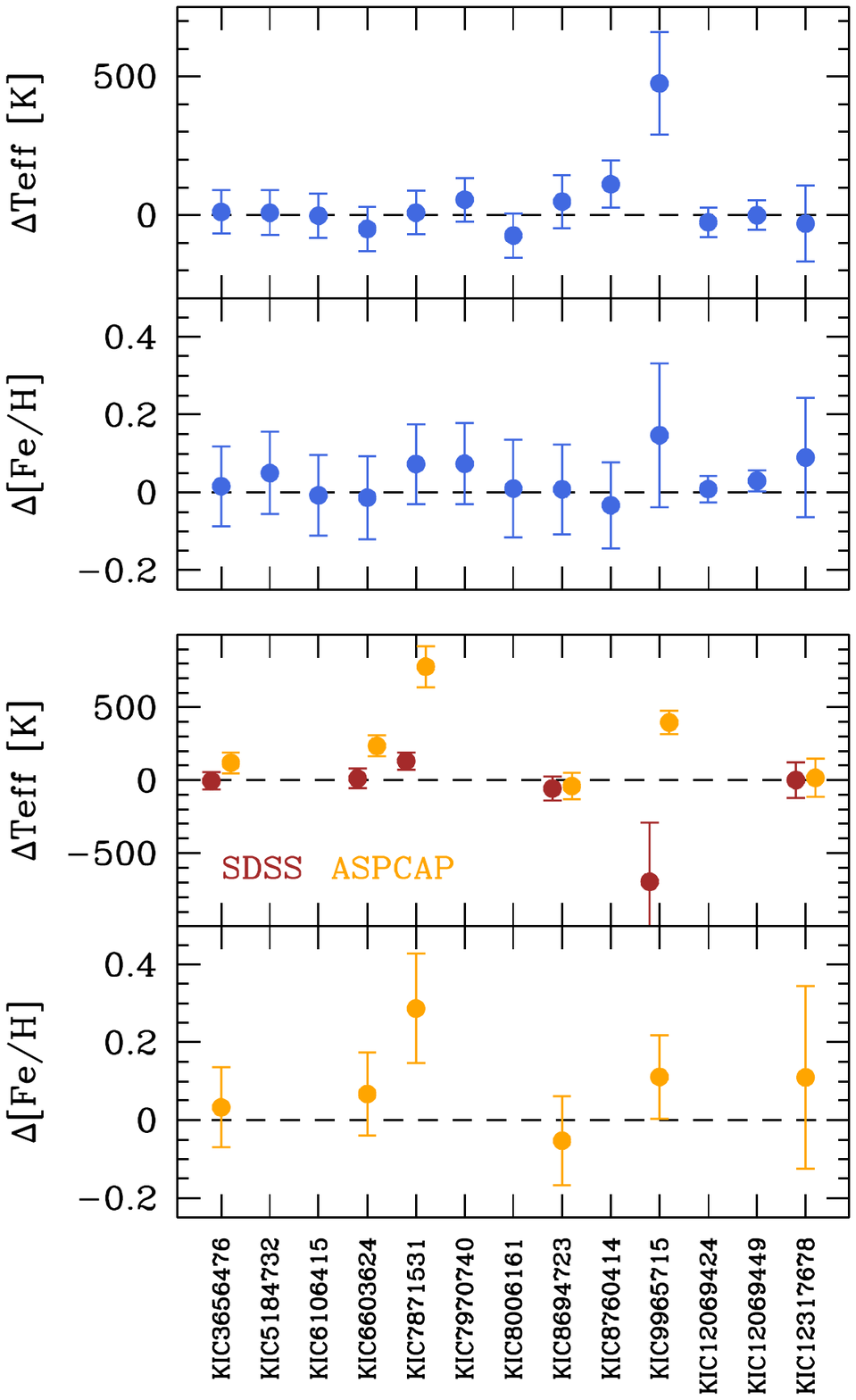}
\caption{{\it Upper panels:} differences (this study minus literature) between our $T_\mathrm{eff}$ and [Fe/H] values and those adopted by C17 and SA17 for their seismic modelling. {\it Bottom panels:} same as upper panels, but for the SDSS and ASPCAP non-seismic constraints adopted by \citet{serenelli17}.}
\label{fig_comparison_with_adopted_values_for_seismic_modelling}
\end{figure}

A comparison with the results of \citet{buchhave15} obtained with the Stellar Parameters Classification tool ({\tt SPC}) is particularly relevant because, with the exception of \object{KIC 9965715} and \object{16 Cyg AB}, their $T_\mathrm{eff}$ and [Fe/H] values were adopted by C17 and SA17 to perform their seismic modelling. A constrained analysis was also enforced by adopting as priors the seismic gravities of \citet{chaplin14}. The results of \citet{buchhave15} are based on data obtained with the TRES \'echelle spectrograph installed on the 1.5-m telescope of the Fred Lawrence Whipple Observatory (Mt. Hopkins, Arizona). The spectra have similar characteristics as ours ($R$ $\sim$ 44,000 and covering 3800-9100 \AA), but the S/N is modest.\footnote{In the range 30-110 per resolution element at 5110 \AA \ based on information about the observing runs in \citet{furlan18}.} Despite this difference in terms of data quality, there is a very close agreement between the stellar parameters without any evidence for systematic discrepancies (Fig.~\ref{fig_comparison_with_adopted_values_for_seismic_modelling}).

In sharp contrast, our $T_\mathrm{eff}$ and [Fe/H] values for \object{KIC 9965715} are discrepant with those of different origin used by C17 and SA17 to model the {\it Kepler} data.\footnote{The adopted values are the preliminary {\it Kepler} Asteroseismology Science Consortium (KASC) estimates available at the time of the analysis (Silva Aguirre, private communication) and are not taken from \citet{pinsonneault12,pinsonneault14}, as erroneously quoted in C17 and SA17.} We find that the star is less metal poor and much hotter, which is much more in line with a F2 V classification and in agreement with the conclusions of \citet{molenda13} and \citet{brewer16}. On the other hand, \citet{compton18} reported problems when trying to model this star with the $T_\mathrm{eff}$ adopted by C17 and SA17.

Overall, one can note a poorer level of agreement between our values and those quoted in the APOCASK\footnote{APOKASC is a joint collaboration between APOGEE and the KASC.} catalogue \citep{serenelli17}, either derived from the analysis of near-IR APOGEE DR13 spectra with the APOGEE Stellar Parameters and Chemical Abundances Pipeline (ASPCAP) or, to a lesser extent, from Sloan Digital Sky Survey (SDSS) $griz$ photometry. The ASPCAP $T_\mathrm{eff}$ scale is indeed known to be too cool for dwarfs and subgiants \citep[e.g.][]{serenelli17,martinez19}, which likely explains the [Fe/H] discrepancy also seen in Fig.~\ref{fig_comparison_with_adopted_values_for_seismic_modelling}. Generally speaking, it is commonplace to find noticeable differences between [Fe/H] determinations based on optical or near-IR spectra, or even between the results of different pipelines applied to APOGEE data \citep[][]{sarmento20}.

We present in Table \ref{tab_parameters_literature} a global comparison with other literature studies. \citet{molenda13} and \citet{furlan18} used various methods. For the former work, there are discrepancies amounting to up to almost 0.5 dex between the spectroscopic and seismic gravities. One can note that a cooler/hotter $T_\mathrm{eff}$ scale is often associated to lower/larger surface gravities on average. This is a well-known consequence of the degeneracy in the determination of these two quantities from spectroscopy \citep[e.g.][]{torres12} that implies that the $T_\mathrm{eff}$ scale is sensitive to the presence of systematic biases in the determination of $\log g$. Compared to our results, the $T_\mathrm{eff}$ and $\log g$ differences indeed appear to be positively correlated for the data of \citet{molenda13} and \citet{brewer16}. This is also true for the {\tt SPC} and combined results of \citet[][]{furlan18}. We find a good agreement with our $T_\mathrm{eff}$ scale when allowance is made for the fact that the spectroscopic gravities for the unconstrained studies above deviate from the very precise seismic estimates. Nonetheless, there is evidence that our metallicities are on average larger by $\sim$0.04 dex than previous studies. Although generally speaking there is a fair consistency between the various sets of parameters, a closer look on a star-to-star basis reveals that the agreement is very good for solar analogues, but considerably worsens for hotter and/or more metal-poor stars. This is not surprising because a robust determination is much more difficult to achieve.

\begin{table*}[h!]
\caption{Comparison with stellar parameters in the literature. The mean differences are this study minus literature. $N$ is the number of stars in common. The $T_\mathrm{eff}$ and [Fe/H] values of \citet{buchhave15} were used to estimate the seismic ages of most of our targets. Note that they provide the mean metallicity, not [Fe/H]. For \citet{molenda13}, we averaged for a given star and method (either {\tt ROTFIT} or {\tt MOOG}) the results obtained with different spectrographs. }
\label{tab_parameters_literature} 
\centering
\begin{tabular}{l|cccr}
\hline\hline                                          
Study and tool                       & $\langle$$\Delta$$T_\mathrm{eff}$$\rangle$ [K] & $\langle$$\Delta$$\log g$$\rangle$ & $\langle$$\Delta$[Fe/H]$\rangle$ & $N$\\
\hline
\citet{bruntt12}\tablefootmark{a}    &                 &                 &                 &  \\
{\tt VWA}                            &   +68$\pm$79    & --0.01$\pm$0.02 &  +0.05$\pm$0.09 &  9\\
\hline
\citet{molenda13}                    &                 &                 &                 &  \\
{\tt ROTFIT}                         &   +85$\pm$72    &  +0.23$\pm$0.12 &  +0.02$\pm$0.12 &  9\\
{\tt MOOG}                           &  --43$\pm$101   & --0.12$\pm$0.20 &  +0.01$\pm$0.05 &  9\\
\hline
\citet{buchhave15}\tablefootmark{a}  &                 &                 &                 &  \\
{\tt SPC}                            &   +9$\pm$54     &  +0.00$\pm$0.01 &  +0.03$\pm$0.05 & 10\\
\hline
\citet{brewer16}\tablefootmark{b}    &                 &                 &                 &  \\
{\tt SME}                            &   +28$\pm$57    &  +0.03$\pm$0.07 &  +0.02$\pm$0.06 & 11\\
\hline
\citet{furlan18}                     &                 &                 &                 &  \\
{\tt SPC}                            &   +36$\pm$114   &  +0.08$\pm$0.18 &  +0.06$\pm$0.08 & 12\\
{\tt Kea}                            &   +43$\pm$73    &  +0.01$\pm$0.12 &  +0.09$\pm$0.09 & 10\\
{\tt SpecMatch}                      &   +36$\pm$73    &  +0.01$\pm$0.06 &  +0.04$\pm$0.13 &  9\\
{\tt Newspec}                        &   +59$\pm$38    &  +0.07$\pm$0.10 & --0.04$\pm$0.09 &  4\\
combined                             &   +52$\pm$63    &  +0.04$\pm$0.08 &  +0.06$\pm$0.08 & 12\\
\hline
\end{tabular}
\tablefoot{
\tablefoottext{a}{Constrained analysis.}
\tablefoottext{b}{Two entries with different IDs are given in the catalogue for \object{KIC 5184732}, \object{KIC 7970740}, and \object{KIC 8006161}: we chose the values based on the spectrum with the highest S/N.}
}
\end{table*}

\subsection{Comparison with interferometric $T_\mathrm{eff}$ scale}\label{sect_validation_parameters_interferometry}

Four stars in our sample have a nearly model-independent $T_\mathrm{eff}$ determined from a combination of absolute flux and limb-darkened CHARA angular diameter measurements: \object{KIC 6106415} and \object{KIC 8006161} \citep{huber12}, as well as \object{16 Cyg AB} \citep{white13}. We do not discuss the CHARA observations of \object{16 Cyg AB} by \citet{boyajian13} because they were carried out without the PAVO beam combiner whose use has been claimed to yield more reliable measurements for stars that are not well resolved \citep[e.g.][]{casagrande14,karovicova18}. The $T_\mathrm{eff}$ scale of \citet{boyajian13} and \citet{white13} appear indeed discrepant, with the former being $\sim$100 K cooler.

To add a comparison point at higher effective temperatures, we analysed the {\it Gaia} benchmark \object{$\beta$ Vir} (F9 V) following exactly the same steps as the main targets. The reference $T_\mathrm{eff}$ (6083$\pm$41 K) and $\log g$ (4.10$\pm$0.02) values are taken from \citet{heiter15}, and primarily rely on the interferometric and seismic observations of \citet{north09} and \citet{carrier05}, respectively. Our analysis is based on a high-quality spectrum (S/N $\sim$ 295) retrieved from the SOPHIE archives. Because no HE spectra are available, we had to rely on a High-Resolution (HR) spectrum with $R$ $\sim$ 75,000. However, our results are expected to be representative of those obtained with HE spectra. To ensure consistency, a HR asteroid spectrum was also necessary. We make use of the co-addition of the 10 best quality exposures in the archives, which results in a mean S/N of 330. We obtain for \object{$\beta$ Vir}: $T_\mathrm{eff}$ = 6145$\pm$35 K and [Fe/H] = +0.16$\pm$0.03. The recommended metallicity value assuming LTE quoted by \citet{jofre14} is [Fe/H] = +0.21$\pm$0.07. Incidentally, an unconstrained analysis yields a $\log g$ lower by 0.01 dex only. The $T_\mathrm{eff}$ and [Fe/H] values are hence very similar.

\begin{figure}[h!]
\centering
\includegraphics[trim=40 285 10 295,clip,width=1.0\hsize]{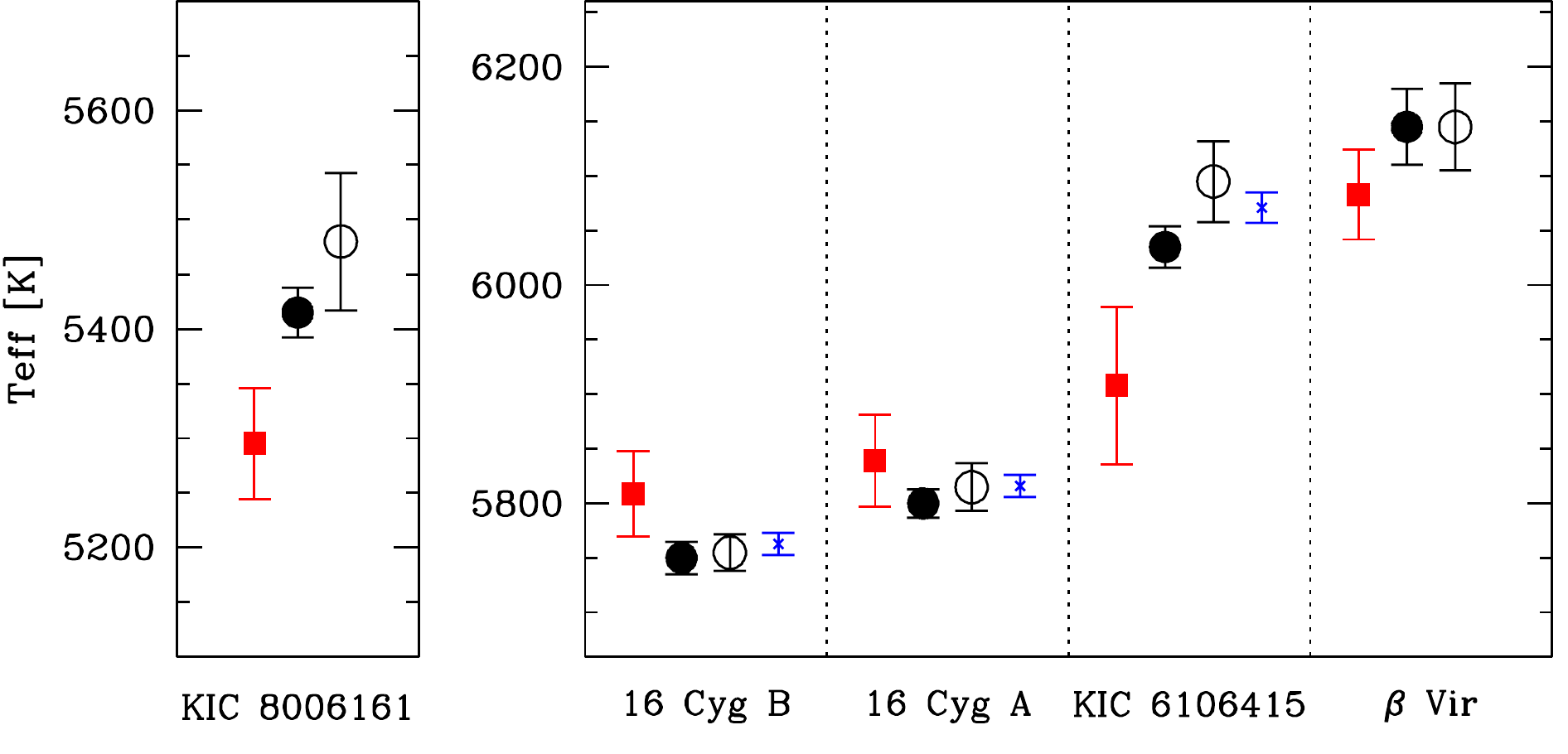}
\caption{Comparison between our ionisation temperatures (constrained and unconstrained shown as filled and open black circles, respectively) and those in the literature derived from interferometric and absolute flux measurements (filled red squares). The blue crosses show the results of N17 determined from iron ionisation balance. The stars are ordered as a function of increasing $T_\mathrm{eff}$.}
\label{fig_comparison_interferometry}
\end{figure}

Figure \ref{fig_comparison_interferometry} shows a comparison between our effective temperatures and those derived from absolute flux and long-baseline interferometric measurements. The values obtained by N17 from iron ionisation balance are also added. Except for \object{16 Cyg AB}, there is indication that the spectroscopic values (either ours or from N17 for \object{KIC 6106415}) are systematically larger. There have been previous reports of systematically hotter $T_\mathrm{eff}$ scales determined from spectroscopy \citep[e.g.][]{boyajian13,brewer16}. Definitive conclusions can hardly be drawn considering the paucity of data points and the recognition that, as discussed above, there is some concern about the robustness of the interferometric data. In particular, we find that the interferometric radii of \object{KIC 6106415} and \object{KIC 8006161} tend to be larger than the seismic estimates of both C17 and SA17. The difference is particularly outstanding for \object{KIC 6106415}, and was ascribed by SA17 to the fact that it is poorly spatially resolved and possibly affected by  calibration problems. The limb-darkened angular diameters, $\theta_\mathrm{LD}$, of \object{16 Cyg AB} are larger than those of \object{KIC 6106415} and \object{KIC 8006161} ($\theta_\mathrm{LD}$ $\sim$ 0.5 vs 0.3 mas), while \object{$\beta$ Vir} is well resolved ($\theta_\mathrm{LD}$ $\sim$ 1.45 mas). Order-of-magnitude estimates show that the $T_\mathrm{eff}$ discrepancy observed in the two stars that are barely resolved may solely be explained by an overestimation of their angular diameters. Reddening should not be an issue given that these stars are all located within 50 pc and comfortably lie inside the Local Bubble. In any case, the cooler $T_\mathrm{eff}$ scale obtained when fixing $\log g$ (see Fig.~\ref{fig_comparison_interferometry}) seems to support the choice made in Sect.~\ref{sect_atmospheric_parameters} of a constrained analysis \citep[see also][]{gent20}.

\section{Validation of chemical abundances}\label{sect_validation_abundances}

The bright, G1.5 V + G3 V binary system \object{16 Cyg AB} is a target of great interest in the context of asteroseismology \citep[e.g.][]{buldgen16,verma14,bazot19,farnir20}, and has been the subject of numerous high-precision abundance studies. Interestingly, there is evidence for a small metallicity difference between the two components that might be attributable to the formation and/or ingestion of planets. The existence of a slight enhancement in metals ($\Delta$[Fe/H] $\sim$ 0.03 dex) in the primary was first convincingly established by \citet{laws01} and is no longer disputed \citep[][and references therein]{tucci_maia19}. Detecting such a subtle difference represents a stringent test of the precision of our results. We compare in Appendix~\ref{appendix_16_cyg_AB} our results to those of N17\footnote{It should be noted that non-LTE corrections were applied to most elements by N17. Unfortunately, their LTE abundances cannot be computed because the mean differential corrections are not provided on a star-to-star basis. However, as discussed by N17, they are exceedingly small for solar analogues and probably irrelevant for the purpose of our discussion. Their uncertainties for elements other than iron also refer to [X/Fe], but we regard them in the following as being representative of those for [X/H].} and \citet[][]{tucci_maia19} that are based on data of exceptional quality ($R$ = 115,000-160,000 and S/N $\sim$ 800). A remarkable agreement is found in both cases, with abundances with respect to hydrogen differing by only $\sim$0.01 dex on average. Furthermore, contrary to previous studies that failed to reveal this peculiarity \citep[][]{schuler11,takeda05}, our results support the slightly more metal-rich nature of the primary. We obtain a weighted mean difference of the abundances with respect to hydrogen of $\sim$0.021 dex, which --- albeit being very small --- is highly significant (see Appendix~\ref{appendix_16_cyg_AB} for further details).
 
\begin{figure}[h!]
\centering
\includegraphics[trim=135 425 100 135,clip,width=1.0\hsize]{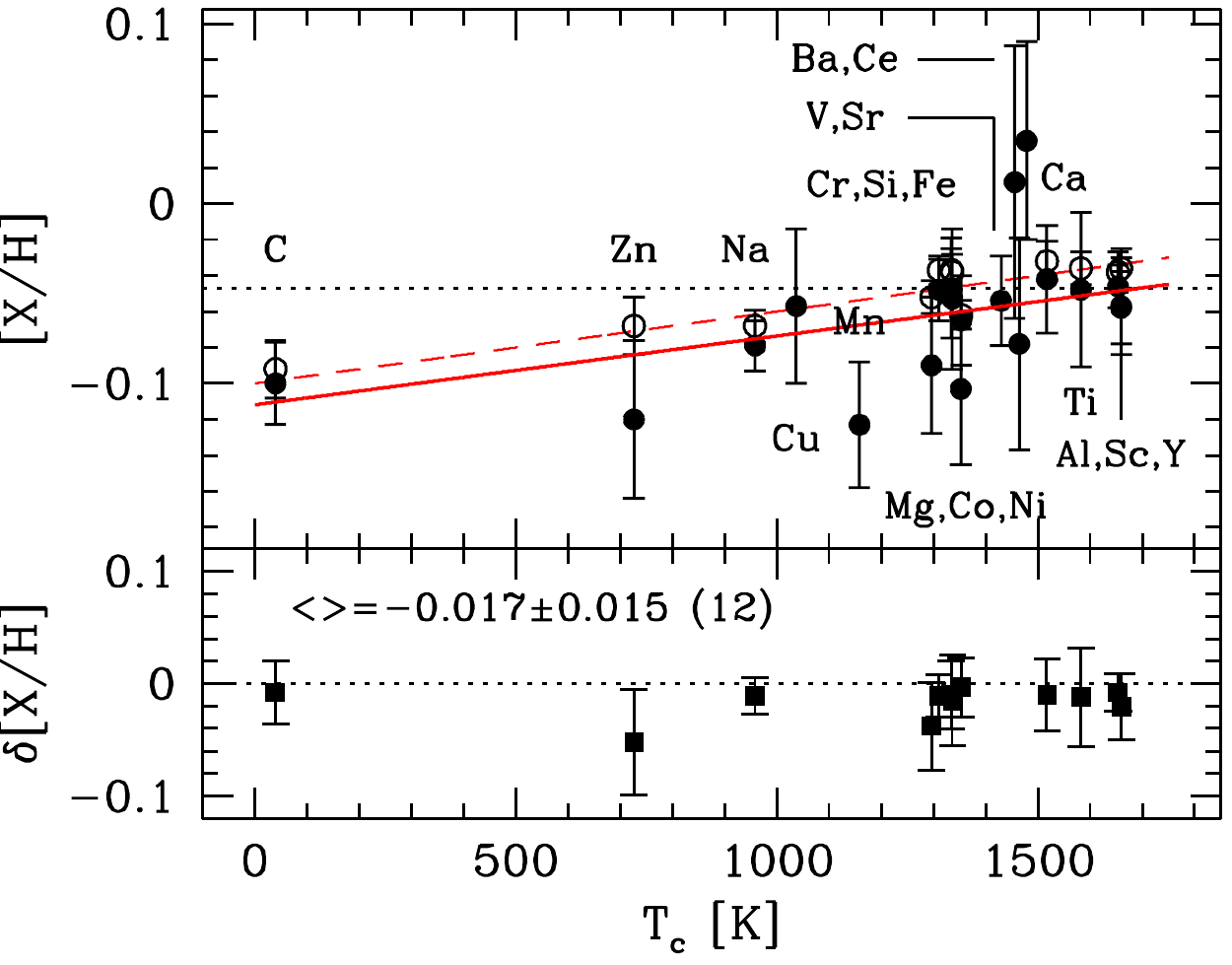}
\caption{{\it Upper panel}: abundance pattern of \object{KIC 6106415} with respect to hydrogen, [X/H], as a function of $T_\mathrm{c}$. Our [X/H] results and those of N17 are shown as filled and open circles, respectively. A dotted, horizontal line is drawn at our [Fe/H] value. The solid line shows our weighted, linear fit of [X/H] as a function of $T_\mathrm{c}$. The fit obtained by N17 is overplotted as a dashed line. {\it Lower panel}: differences (our values minus theirs) with respect to N17, $\delta$[X/H]. The average $\delta$[X/H] value is given (the number of elements in common is indicated in brackets). To guide the eye, a dotted line is drawn at $\delta$[X/H] = 0.}
\label{fig_condensation_KIC06106415}
\end{figure}

The G0 V star \object{KIC 6106415} was studied by N17. We find an ionisation $T_\mathrm{eff}$ lower by $\sim$35 K. A comparison between the abundance data is shown in Fig.~\ref{fig_condensation_KIC06106415}, as a function of the 50\% condensation temperature for a solar-system composition gas \citep[$T_\mathrm{c}$;][]{lodders03}. Although there is a slight, systematic offset ($\sim$0.02 dex) between the [X/H] values, the scatter is very small ($\sim$0.015 dex). The zero-point offset affecting the absolute abundances may have several causes (e.g. continuum placement). The largest discrepancy ($\sim$0.05 dex) is found for Zn, which has uncertain abundances in both studies \citep[see also][]{nissen20}. The agreement is remarkable when the abundances with respect to iron are considered: $\langle$$\Delta$[X/Fe]$\rangle$ (this study minus N17) = --0.007$\pm$0.015 dex.

Although the tests described above suggest that our abundance results are precise at the 0.02-0.03 dex level for early G dwarfs, we warn the reader that this conclusion \emph{cannot} be extended to the stars in our sample with properties significantly departing from solar. It is because the effects of various physical phenomena (departures from LTE, convection inhomogeneities, atomic diffusion) no longer cancel out to first order through our differential analysis.

\begin{figure*}[h!]
\centering
\includegraphics[trim=50 325 10 130,clip,width=0.7\hsize]{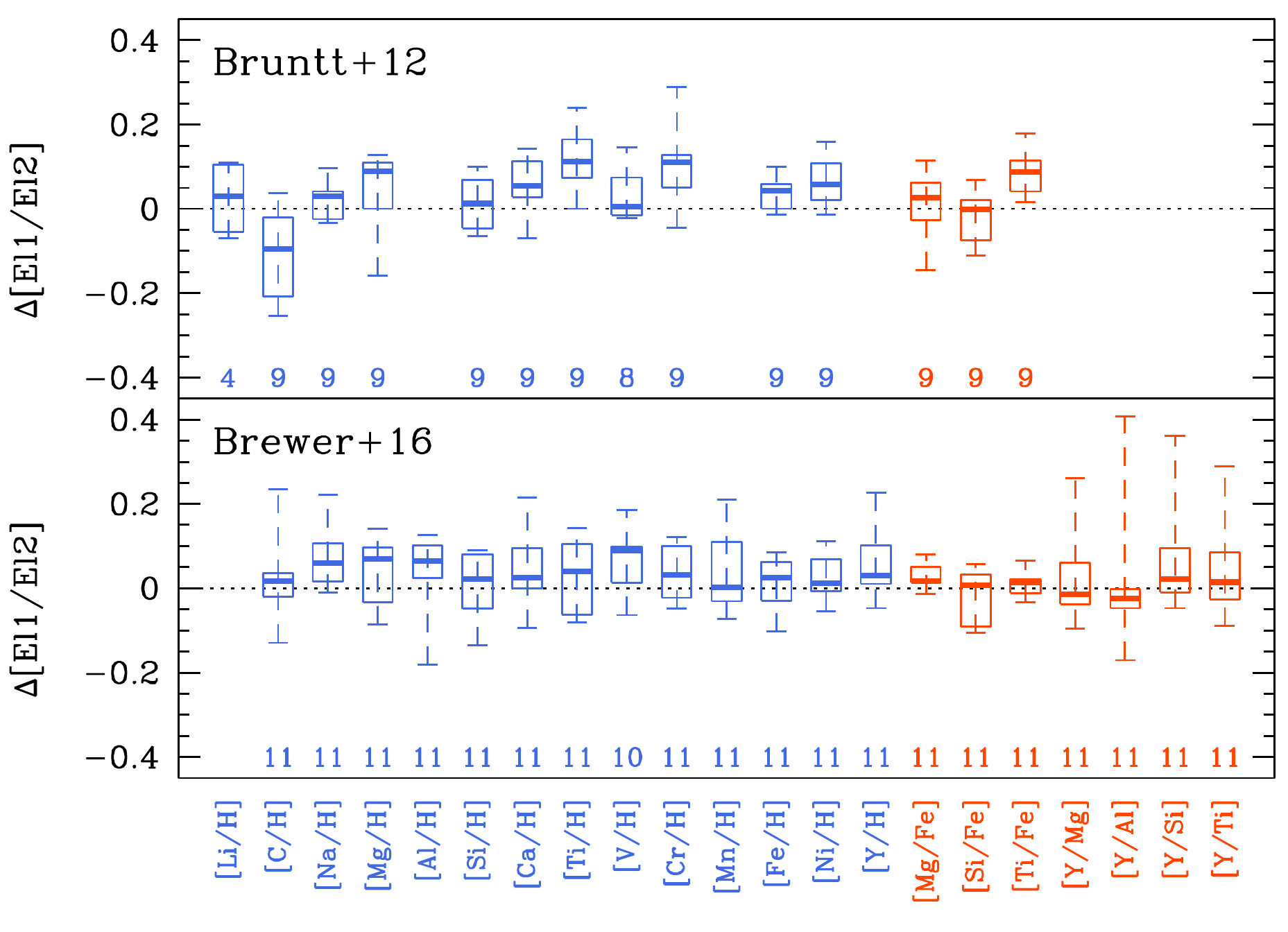}
\caption{Comparison between our abundance ratios and those of \citet[][]{bruntt12} and \citet[][]{brewer16}. The abundance differences are this study minus literature. The vertical dashed line connects the extreme values. The box covers the first to third quartile of the data, while the thick horizontal line inside the box shows the median. As recommended by \citet{bruntt12}, we chose their \ion{Fe}{ii}-based abundances for [Fe/H] (but used the \ion{Mg}{i}, \ion{Si}{i}, \ion{Ti}{i}, and \ion{Cr}{i} abundances). For \object{KIC 5184732}, \object{KIC 7970740}, and \object{KIC 8006161}, we chose the values of \citet[][]{brewer16} corresponding to the spectrum with the highest S/N. The number of stars the calculation is based on is indicated for each abundance ratio. The red symbols show the abundance ratios used as age indicators (Sect.~\ref{sect_discussion}).}
\label{fig_comparison_abundances}
\end{figure*}

For a more general validation of the elemental abundances, we consider the comprehensive studies of the {\it Kepler} targets undertaken by \citet{bruntt12} and \citet{brewer16}. The surface gravity was held fixed to the seismic value in the former work. Both analyses are based on spectral synthesis performed with (semi-)automatic tools. The results are compared to ours in Fig.~\ref{fig_comparison_abundances}. Although some outliers are evident (e.g. \object{KIC 12317678} when compared to \citealt{brewer16}), there is a reasonable agreement with only a slight offset in general for the abundance ratios with respect to hydrogen (of the order of 0.04 dex on average when considering the median differences). The most conspicuous difference is the lower carbon abundances we obtain compared to \citet{bruntt12}, but their uncertainties are quite large for some stars (up to 0.15 dex). However, whether or not we exclude this element, we tend to find a slightly better agreement with \citet{brewer16}.

\section{Discussion}\label{sect_discussion}

\subsection{Age inferences from abundance data for the LEGACY sample}\label{sect_discussion_age_legacy}

Several linear relations directly linking isochrone ages and abundance ratios have been proposed for solar twins/analogues \citep[e.g.][]{bedell18,nissen15,tucci_maia16}. Such linear relationships may be an oversimplification \cite[][]{spina16,spina18}. It is also becoming increasingly clear that other variables must be taken into account in order to apply them to stars with parameters strongly departing from the solar values \citep[e.g.][]{feltzing17}. Recently, \citet[][hereafter DM19]{delgado_mena19} used a large sample of main-sequence FGK stars observed as part of the HARPS-Guaranteed Time Observations (HARPS-GTO) program to empirically explore the dependency of the abundance-isochrone age relationships as a function of $T_\mathrm{eff}$, metallicity, and mass. They found that several abundance ratios, when combined with one or two stellar parameters, can be used to infer stellar ages to within 1.5-2 Gyrs. They proposed three types of relationships: 1D (age vs abundance ratio), 2D (age vs abundance ratio and either $T_\mathrm{eff}$, [Fe/H], or $M$) and finally 3D (age vs abundance ratio and two of either $T_\mathrm{eff}$, [Fe/H], and $M$). The main rationale behind including a $T_\mathrm{eff}$ and $M$ dependency is to capture the effect of atomic diffusion, although it can be noted that using metal abundance ratios largely reduces its importance \citep{dotter17}. They concluded that 2D relations perform better compared to those in 1D (especially when [Fe/H] or $M$ are folded in) and that 3D relations do not lead to a major additional improvement. Furthermore, it was found that 3D relations yield similar results regardless of the choice of the two stellar parameters. It should be noted that, contrary to some studies \citep[e.g.][]{nissen16}, thin- and thick-disc stars were not separated when constructing these calibrations. Thick-disc stars constitute about 20\% of the subsample of 354 stars with the most precise isochrone ages (uncertainty below 1.5 Gyr) built for that purpose. Therefore, we apply the calibrations to our whole sample in the following. We discuss below the ages we obtain from these relationships and compare them to the seismic estimates.

\begin{figure}[h!]
\centering
\includegraphics[trim=20 280 15 150,clip,width=1.0\hsize]{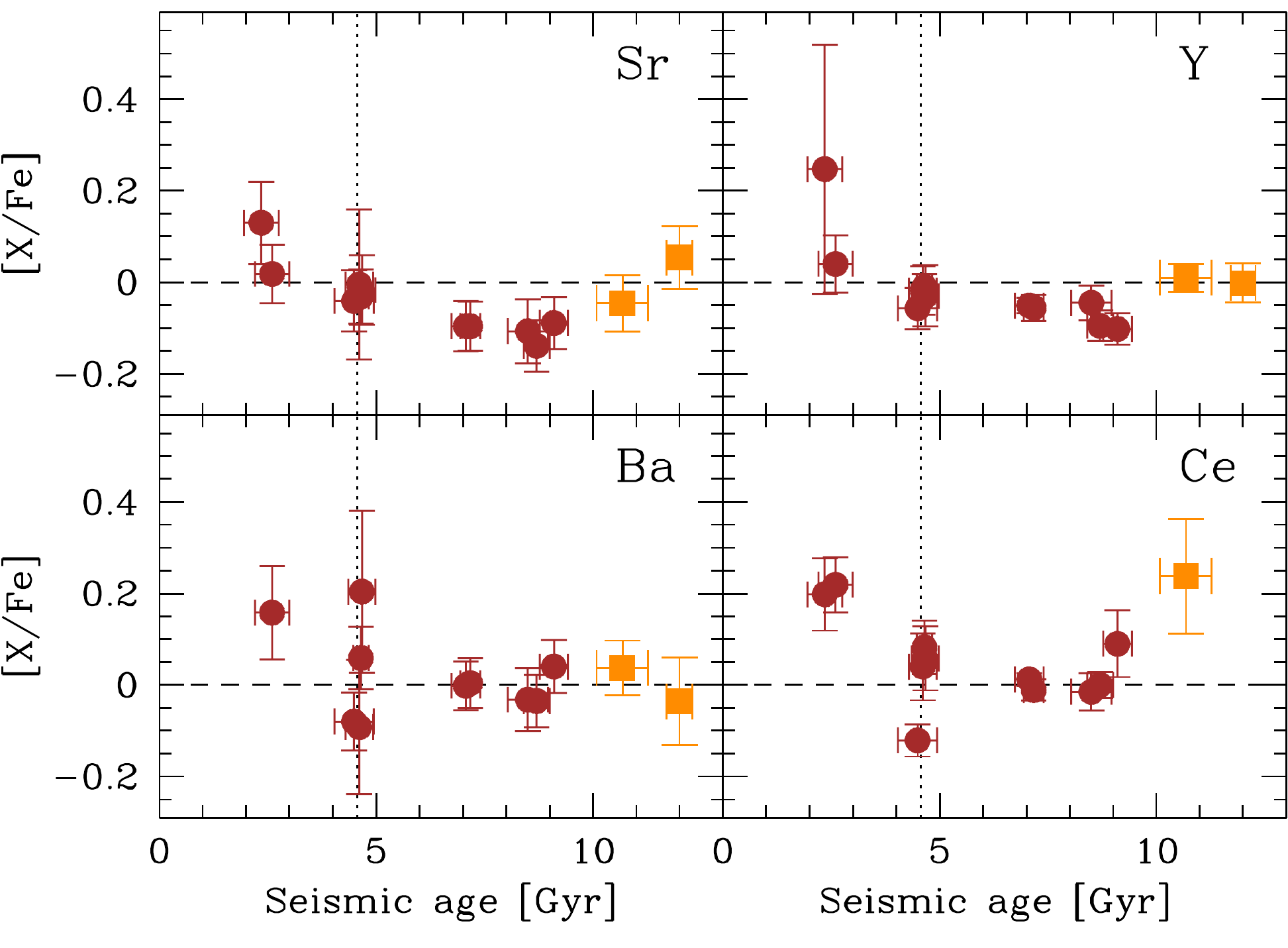}
\caption{Abundances of the neutron-capture elements relative to iron as a function of seismic age. The thin- and thick-disc stars are shown as circles and squares, respectively. The solar age is shown as a vertical dotted line.}
\label{fig_abundances_vs_age_neutron_process}
\end{figure}

DM19 proposed several abundance ratios that can be used as ``chemical clocks'', among which fourteen can be computed from our data. Although we refrain from discussing the time evolution of other abundance ratios in our small dataset, let us briefly comment on the behaviour of the predominantly slow ($s$-) neutron-capture elements Sr and Y. As previously found \citep[e.g.][]{spina18}, a smooth decline of [Sr/Fe] and [Y/Fe] with increasing age is seen. However, there is some evidence for an upturn for the two thick-disc stars with an age above $\sim$10 Gyrs (Fig.~\ref{fig_abundances_vs_age_neutron_process}). This is consistent with a picture in which there was  a more vigorous production during the formation of the thick disc of $s$-elements from low-mass asymptotic giant branch (AGB) stars relative to iron from Type Ia supernovae \citep[e.g.][]{battistini16}. This is apparently not seen in Fig.~\ref{fig_abundances_vs_age_neutron_process} for Ba, while the only Ce abundance at old ages is very uncertain. See DM19 for a discussion of the behaviours of the light and heavy $s$-elements.

\begin{sidewaysfigure*}[p]
\begin{minipage}[h!]{0.33\textwidth}
\centering
\includegraphics[trim=50 215 145 213,clip,width=1\hsize]{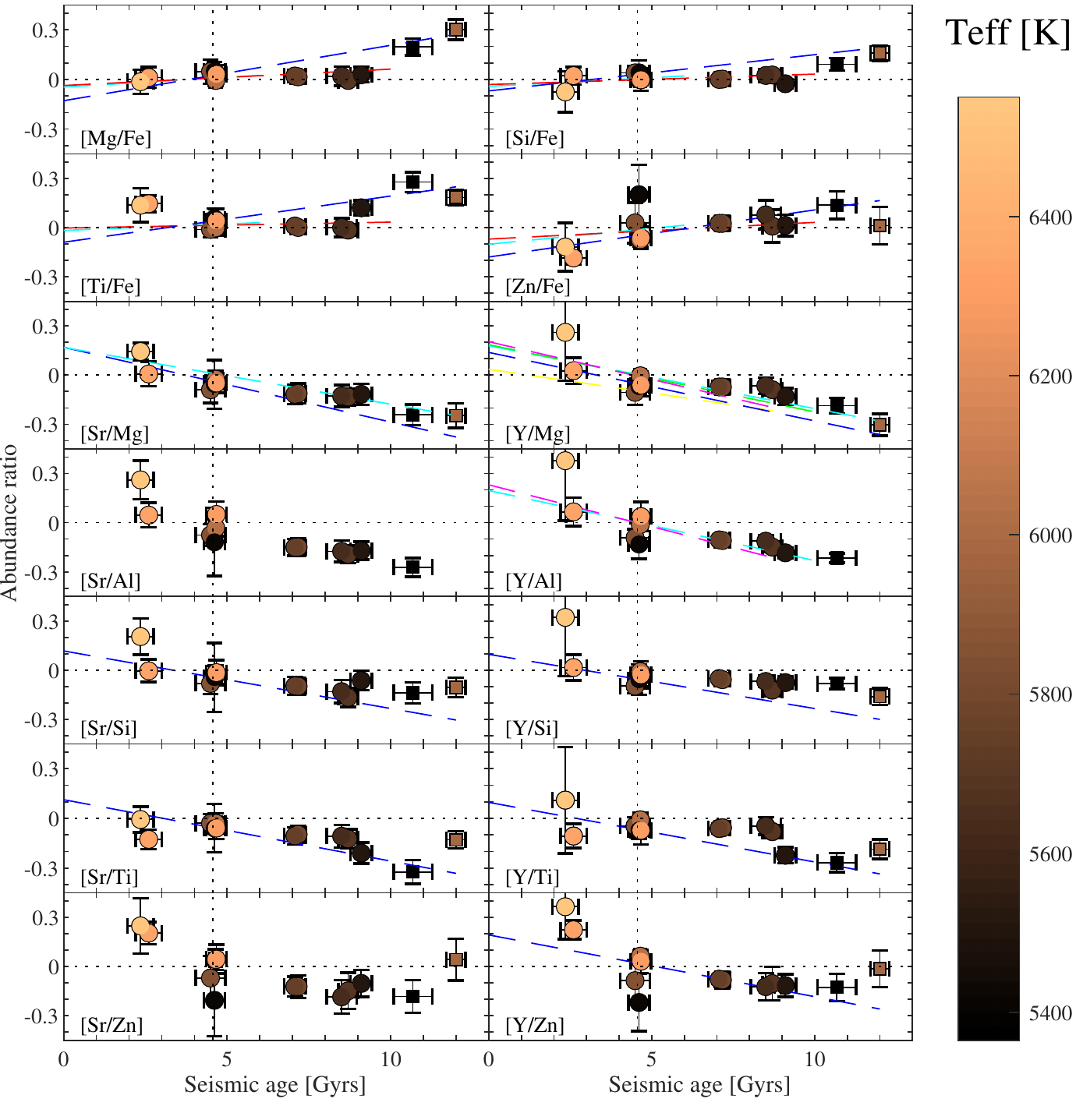}
\end{minipage}
\begin{minipage}[h!]{0.33\textwidth}
\centering
\includegraphics[trim=50 215 145 213,clip,width=1\hsize]{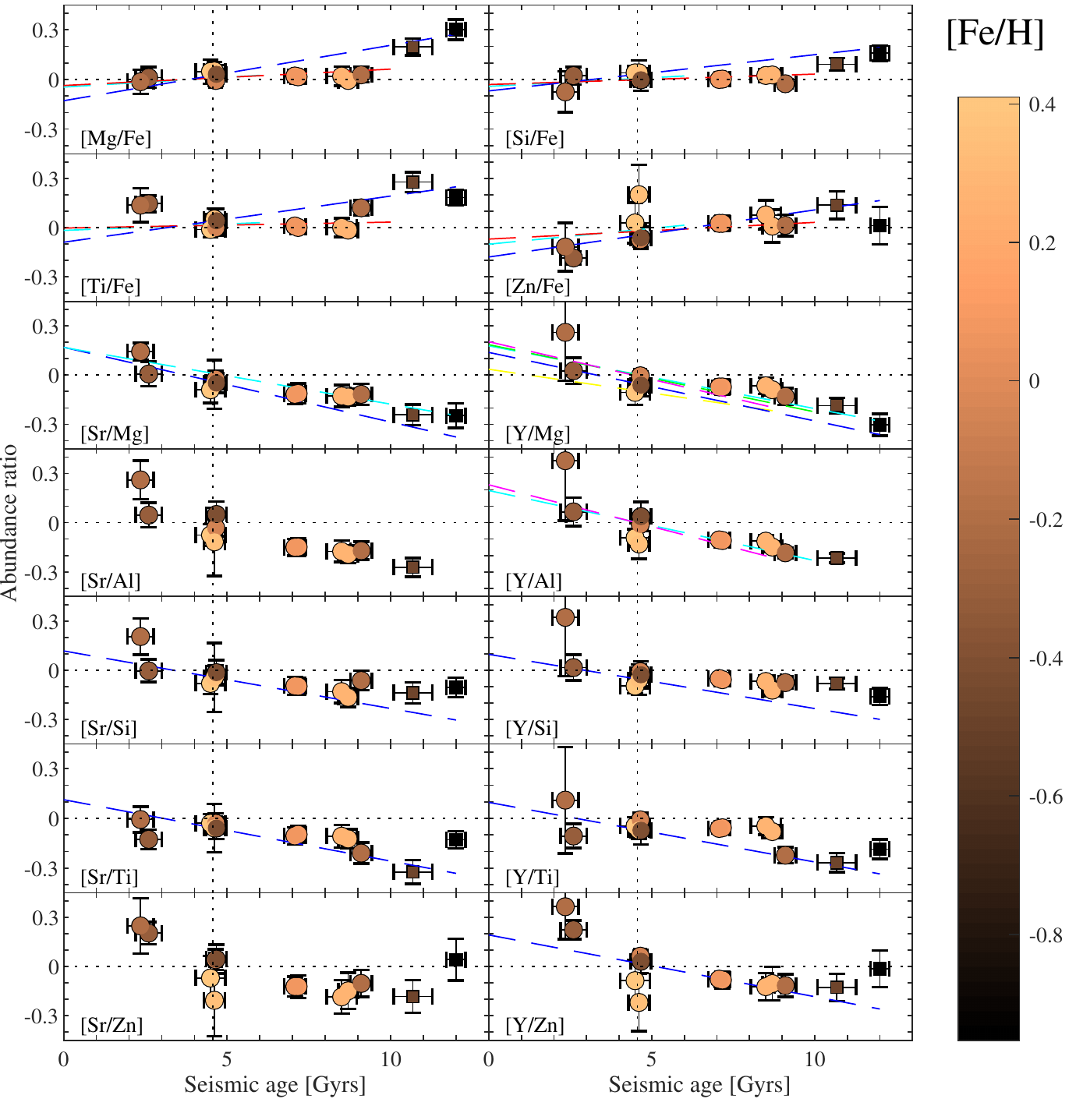}
\end{minipage}
\begin{minipage}[h!]{0.33\textwidth}
\centering
\includegraphics[trim=50 215 145 213,clip,width=1\hsize]{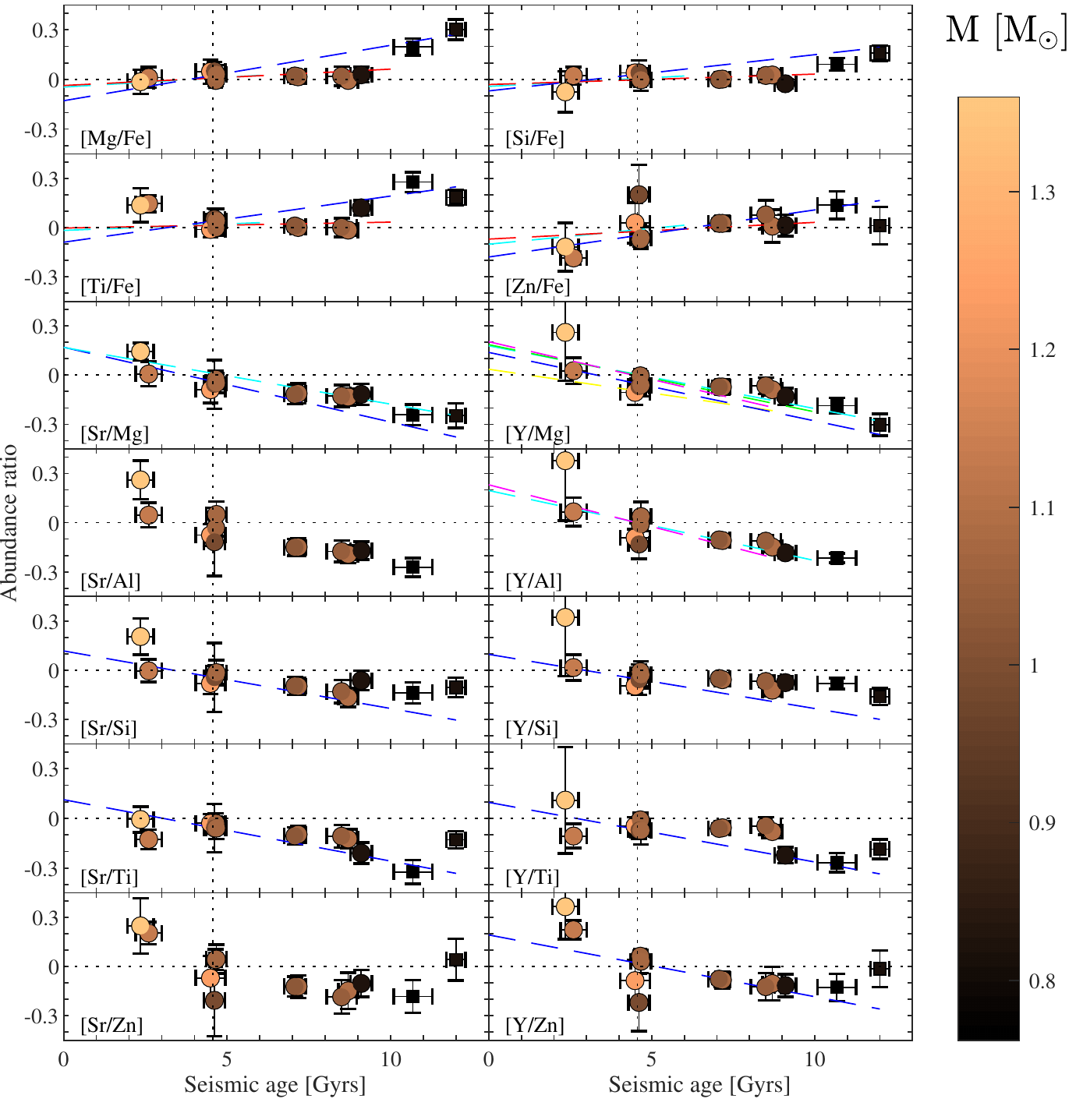}
\end{minipage}
\caption{{\it Left panels}: variation of the abundance indicators, as a function of seismic age. The data are colour coded as a function of $T_\mathrm{eff}$. The thin- and thick-disc stars are shown as circles and squares, respectively. The solar age is shown as a vertical dotted line. {\it Middle and right panels}: same as left panels, but colour coded as a function of [Fe/H] and $M$, respectively. Linear 1D relations in the literature linking abundance ratios and ages are overplotted as dashed lines: \citet[][{\it red}]{bedell18}, DM19 ({\it blue}), \citet{nissen16} and \citet[][{\it cyan}]{nissen20}, \citet[][{\it magenta}]{spina18}, \citet[][{\it yellow}]{titarenko19}, and \citet[][{\it green}]{tucci_maia16}.}
\label{fig_abundances_vs_seismic_age}
\end{sidewaysfigure*}

We show in Fig.~\ref{fig_abundances_vs_seismic_age} how some abundance ratios selected by DM19 vary as a function of seismic age. The data are colour coded as a function of $T_\mathrm{eff}$, [Fe/H], and $M$. Previous 1D relations proposed in the literature are also overplotted \citep[DM19;][]{nissen16,nissen20,tucci_maia16,spina18,bedell18,titarenko19}. We recall that, with the exception of DM19 and \citet{titarenko19}, these relationships only apply to solar analogues and are not expected to provide a good fit to our data. Also, some of them may not be valid for old or thick-disc stars. The deviations with respect to 1D calibrations may be ascribed to stellar parameters differing from the solar values. A good example is [Y/Si] where the systematic deviations clearly noticeable for the old stars are largely removed when using 2D or 3D relations. A more general discussion is provided below.

We compare in Table \ref{tab_ages} for the whole sample the seismic ages and those obtained for each type of relation (e.g. 1D) and abundance indicator (e.g. [Y/Mg]). Both the uncertainties in the abundance ratios and in the calibrations were propagated to the ages. The best abundance indicators proposed by DM19 differ depending on the parametrisation. In this respect, it is interesting to note that [Y/Al], which was often discussed in the literature in the context of solar analogues \citep[e.g.][]{nissen16}, is not among the ratios displaying the best 1D relation when considering the much more heterogeneous sample of DM19.

The young, F-star \object{KIC 12317678} falls close to the upper $T_\mathrm{eff}$ boundary of the relationships of DM19 and has abundance ratios that are particularly uncertain. However, we find that including it in our sample does not bias our results. For instance, removing it leads to mean deviations between the seismic- and abundance-based ages for the whole sample that differ very little (in the range $\pm$0.2 Gyr only depending on the type of relation). This is mainly because unphysical (negative) ages are often obtained --- especially when they are based on the high Y abundance --- and evidently ignored. Negative ages were also obtained in two cases for another young star, \object{KIC 9965715}.

As seen in Table \ref{tab_ages}, the standard deviation of the distributions of age differences is typically $\sim$1.5-2 Gyrs, which is comparable to the intrinsic precision of the calibrations of DM19. To put these results in perspective, internal errors of the order of 0.5-1 Gyr are common for solar analogues \citep[e.g.][]{tucci_maia16,spina18}. Generally speaking, determining ages from a single abundance indicator, whatever the relation used, is not recommended because it is prone to errors amounting to up to several Gyrs. Some abundance-based ages (e.g. using [Ti/Fe]) appear to be less precise than others. There is tentative evidence that the age scatter decreases as the intrinsic quality of the 1D and 2D $T_\mathrm{eff}$ relations increases (as parametrised by the goodness-of-fit measure, adj-$R^2$; see DM19 for definition). It is not seen for the other relations, but the adj-$R^2$ range is much smaller. The scatter between the abundance- and seismic-based ages is reduced by typically $\sim$20\% when averaging the results of all abundance ratios. As may be expected, combining the results of many chemical clocks leads to more precise ages, but the gain is relatively modest.

Ages estimated from pair of elements with significantly different condensation temperatures, $T_c$, may be more sensitive to star-to-star differences in their abundance-$T_c$ trend \citep[N17;][]{nissen18}. Namely, stars with the same age might have different volatile-to-refractory abundance ratios \citep[e.g.][]{biazzo15}. In our case, one might thus expect the two abundance indicators involving the volatile element Zn ([Sr/Zn] and [Y/Zn]) to yield less precise ages. However, it is not borne out by our results (Table \ref{tab_ages}). 

\begin{table*}[h!]
  \caption{Unweighted mean age deviation with respect to seismic estimates (abundance-based minus seismic) for each abundance indicator and relation (the number in brackets is the number of stars the calculation is based on). The bottom row provides the grand mean for a given relation. A blank indicates that the abundance ratio is not among the best age indicators for the relevant relation according to DM19.}
\label{tab_ages} 
\hspace*{-0.2cm}
\begin{tabular}{l|r|rrr|r}
  \hline\hline
Abundance indicator   & \multicolumn{1}{c}{1D}    & \multicolumn{3}{c}{2D}                                                                     & \multicolumn{1}{c}{3D}\\
            &                           & \multicolumn{1}{c}{$T_\mathrm{eff}$} & \multicolumn{1}{c}{[Fe/H]} & \multicolumn{1}{c|}{$M$} & \multicolumn{1}{c}{$T_\mathrm{eff}$ and [Fe/H]}\\
\hline
$[$Mg/Fe$]$ & --1.216$\pm$2.324 (13)    & \multicolumn{1}{c}{...} & \multicolumn{1}{c}{...} & \multicolumn{1}{c|}{...} & \multicolumn{1}{|c}{...} \\
$[$Si/Fe$]$ & --2.419$\pm$2.492 (12)    & --1.554$\pm$2.041 (12)  & \multicolumn{1}{c}{...} & \multicolumn{1}{c|}{...} & \multicolumn{1}{|c}{...} \\
$[$Ti/Fe$]$ & --0.917$\pm$3.727 (13)    & --0.760$\pm$2.924 (13)  & \multicolumn{1}{c}{...} & \multicolumn{1}{c|}{...} & \multicolumn{1}{|c}{...} \\
$[$Zn/Fe$]$ &   0.051$\pm$3.337 (12)    & \multicolumn{1}{c}{...} & --0.503$\pm$2.209 (13)  &   0.012$\pm$2.391 (13)   & \multicolumn{1}{|c}{...} \\
\hline
$[$Sr/Mg$]$ & --0.968$\pm$1.359 (13)    & --0.536$\pm$1.381 (13)  & --1.082$\pm$1.963 (13)  & --0.283$\pm$1.219 (12)   & --0.845$\pm$1.667 (13)    \\
$[$Sr/Al$]$ & \multicolumn{1}{|c|}{...} & \multicolumn{1}{c}{...} & --0.789$\pm$2.350 (13)  & \multicolumn{1}{c|}{...} & \multicolumn{1}{|c}{...} \\
$[$Sr/Si$]$ & --1.375$\pm$2.008 (12)    & \multicolumn{1}{c}{...} & \multicolumn{1}{c}{...} & --0.377$\pm$1.163 (12)   & --0.994$\pm$1.532 (12)    \\
$[$Sr/Ti$]$ & --0.697$\pm$2.186 (13)    & --0.334$\pm$2.089 (13)  & --0.860$\pm$2.556 (13)  & --0.317$\pm$1.935 (13)   & --0.633$\pm$2.132 (13)    \\
$[$Sr/Zn$]$ & \multicolumn{1}{|c|}{...} & \multicolumn{1}{c}{...} & --0.668$\pm$1.713 (13)  & \multicolumn{1}{c|}{...} & \multicolumn{1}{|c}{...} \\
\hline
$[$Y/Mg$]$  & --1.449$\pm$1.603  (12)   & --0.809$\pm$1.487 (12)  & --1.514$\pm$1.826 (12)  & --0.732$\pm$1.407 (12)   & --0.613$\pm$1.497 (12)    \\
$[$Y/Al$]$  & \multicolumn{1}{|c|}{...} & \multicolumn{1}{c}{...} & --1.012$\pm$2.086 (12)  & \multicolumn{1}{c|}{...} & --0.676$\pm$1.934 (13)    \\
$[$Y/Si$]$  & --2.104$\pm$1.955  (12)   & --0.927$\pm$2.083 (12)  & \multicolumn{1}{c}{...} & --0.841$\pm$1.282 (12)   & --0.411$\pm$1.358 (12)    \\
$[$Y/Ti$]$  & --1.507$\pm$2.207  (12)   & --0.464$\pm$2.209 (12)  & \multicolumn{1}{c}{...} & --0.693$\pm$2.061 (12)   & --0.441$\pm$2.017 (12)    \\
$[$Y/Zn$]$  & --0.243$\pm$3.132  (11)   &   1.254$\pm$3.231 (12)  & --0.903$\pm$1.655 (12)  &   0.074$\pm$2.080 (12)   & --0.238$\pm$1.705 (12)    \\
\hline
Mean        & \multicolumn{1}{c|}{--1.028$\pm$1.871}         & \multicolumn{1}{c}{--0.465$\pm$1.720}       & \multicolumn{1}{c}{--0.862$\pm$1.790}       & \multicolumn{1}{c}{--0.466$\pm$1.461}        & \multicolumn{1}{|c}{--0.623$\pm$1.550}          \\
\hline
\end{tabular}
\end{table*}

\begin{figure}[h!]
  \centering
  \includegraphics[trim=150 430 135 150,clip,width=1.0\hsize]{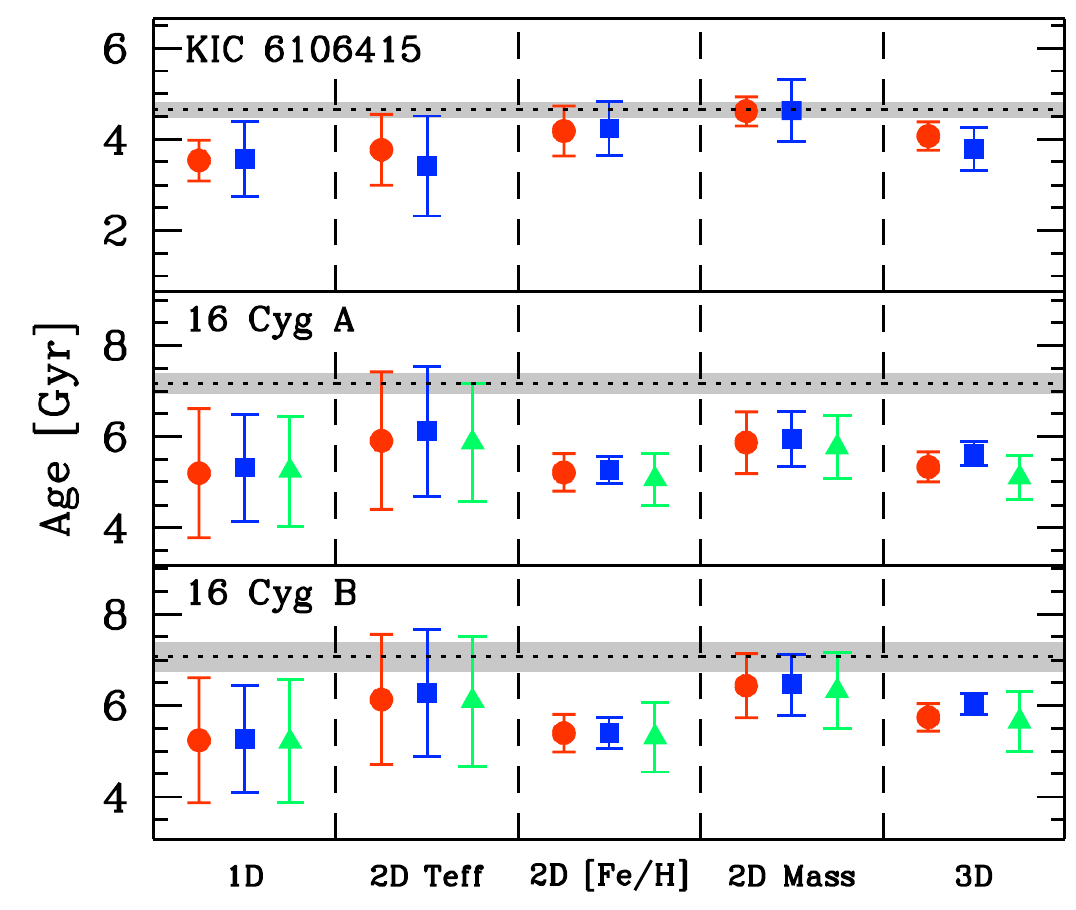}
\caption{Mean ages of \object{KIC 6106415} and \object{16 Cyg AB} obtained using our abundance data ({\it circles}), those of N17 ({\it squares}), and those of \citet[][{\it triangles}]{tucci_maia19}. The shaded horizontal stripes show the seismic ages, along with their $\pm$1$\sigma$ uncertainty.}
\label{fig_comparison_DM19_mean_ages_N17_TM19}
\end{figure}

There is clear evidence that the calibrations provide systematically younger ages compared to asteroseismology by $\sim$0.7 Gyr on average irrespective of the relationship used. To assess the robustness of this result, we computed the ages of \object{KIC 6106415} and \object{16 Cyg AB} using the abundance data of N17 and \citet{tucci_maia19}. As seen in Fig.~\ref{fig_comparison_DM19_mean_ages_N17_TM19}, the discrepancy persists. As an additional test, we redetermined the abundances of the metal-poor, G0 subgiant \object{KIC 8694723} adopting the line lists used to construct the calibrations of DM19. They are taken from \citet{neves09} for Mg, Al, Si and Ti (with a few lines excluded, as described in \citealt{adibekyan12}) and from \citet{delgado_mena17} for Zn, Sr and Y. We do not choose a solar analogue because non-LTE/3D effects may be efficiently erased in that case through a differential analysis whatever the line list. On the contrary, the combined effect might be quantitatively different depending on the choice of the  diagnostic lines when the reference star occupies another region of the parameter space. We obtain ages on average $\sim$0.8 Gyr older. The fact that it apparently solves the discrepancy mentioned above might be accidental: much more extensive tests are needed. We have also used our atmospheric parameters, while systematic differences (e.g. in the $T_\mathrm{eff}$ scale) are also likely to play a role. In any case, it shows that age deviations of the right order of magnitude might be ascribed to small ($\lesssim$ 0.05 dex) zero-point abundance offsets. It is particularly true for Sr and Y because these two elements enter most calibrations. However, systematic differences between various sets of theoretical isochrones are well known. DM19 made use of PARSEC isochrones \citep{bressan12} to infer the ages through Bayesian inference. Adopting another set of evolutionary models may be enlightening in this respect.
 
\begin{figure*}[h!]
\centering
\includegraphics[trim=25 170 20 160,clip,width=1.0\hsize]{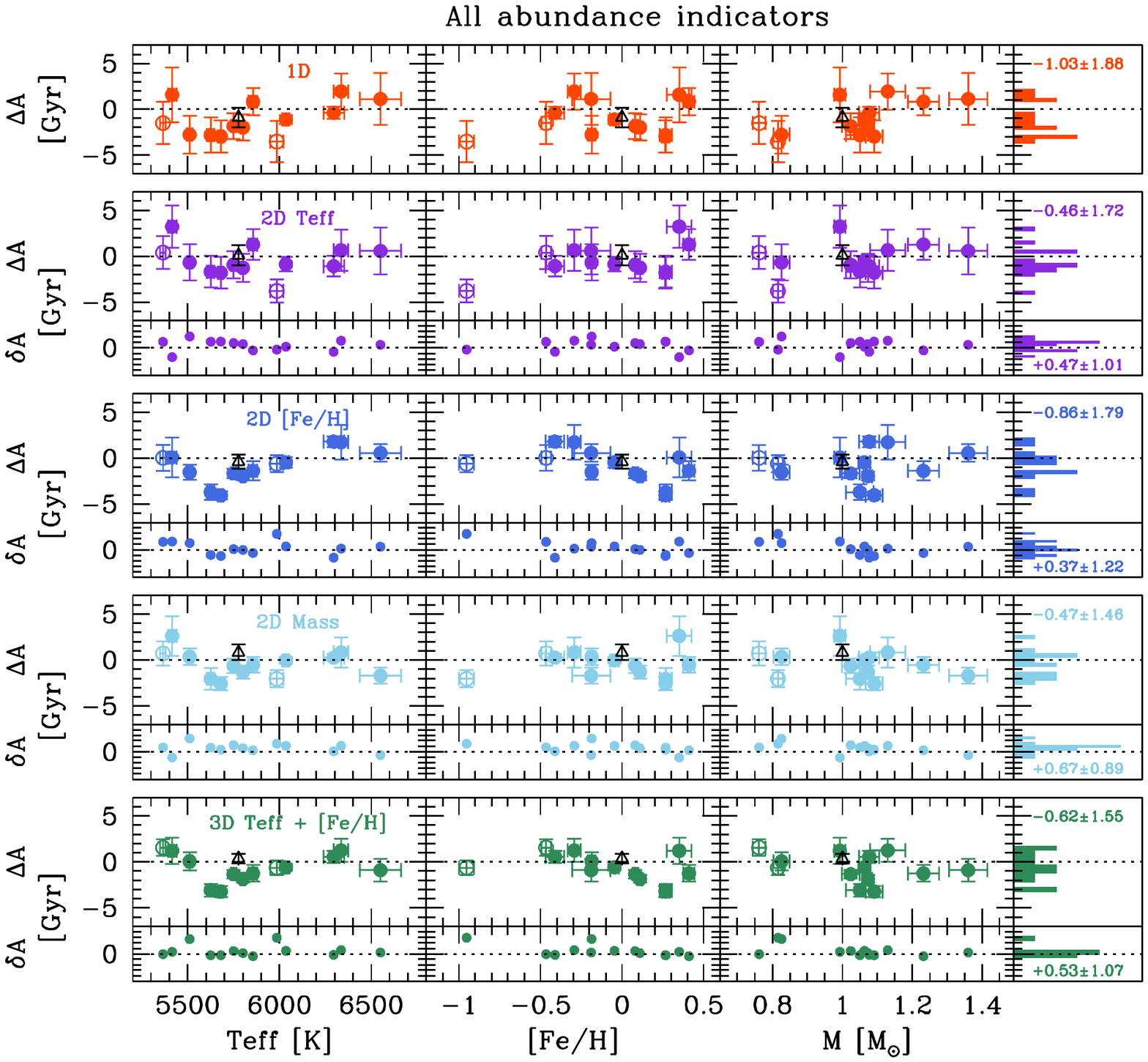}
\caption{Mean age deviation, $\Delta A$, with respect to the seismic estimate (abundance-based minus seismic) when considering all the abundance indicators, as a function of $T_\mathrm{eff}$, [Fe/H], and $M$. The quantity $\delta A$ is a measure of the improvement brought about by using 2D or 3D instead of 1D relations (see Sect.~\ref{sect_discussion_age_legacy}). The thin- and thick-disc stars are shown as filled and open circles, respectively. The Sun is shown as an open triangle. The rightmost panels show the breakdown of the $\Delta A$ and $\delta A$ values.}
\label{fig_comparison_DM19_mean_ages_vs_parameters}
\end{figure*}

\begin{figure*}[h!]
\centering
\includegraphics[trim=25 170 20 160,clip,width=1.0\hsize]{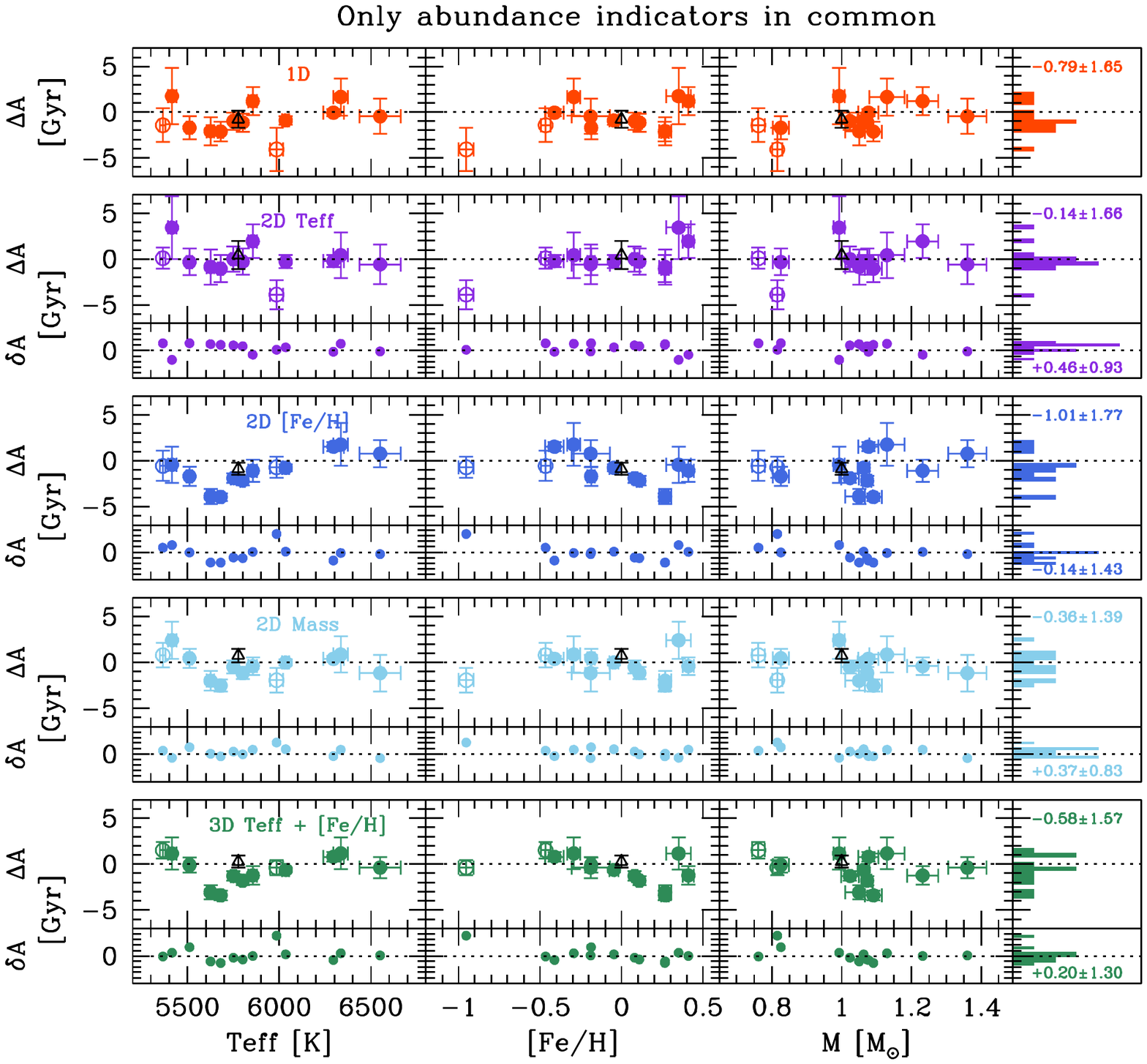}
\caption{Same as Fig.~\ref{fig_comparison_DM19_mean_ages_vs_parameters}, but only for the four abundance indicators common to all the relations.}
\label{fig_comparison_DM19_mean_ages_common_vs_parameters}
\end{figure*}

Figure \ref{fig_comparison_DM19_mean_ages_vs_parameters} illustrates the age deviations, $\Delta A$, as a function of $T_\mathrm{eff}$, [Fe/H], and $M$. The Sun is overplotted to ensure that its age is correctly reproduced and that no systematic age offsets are present. The size of the error bar in this particular case also provides an indication of the age uncertainty arising from the calibrations alone. There is no evidence in Fig.~\ref{fig_comparison_DM19_mean_ages_vs_parameters} that the thick-disc stars are outliers. We define a quantity, $\delta A$, which is a measure of the improvement brought about by using 2D or 3D relations instead of 1D ones. A positive value implies that the age is closer by $\delta A$ Gyrs to the reference seismic value when using a 2D or 3D parametrisation (and vice versa). The relative performance of the various relations is better appraised when examining the $\delta A$ values for the four abundance indicators in common to all relations: [Sr/Mg], [Sr/Ti], [Y/Mg], and [Y/Zn] (Fig.~\ref{fig_comparison_DM19_mean_ages_common_vs_parameters}). The use of more sophisticated relations generally leads to a better agreement with the seismic ages ($\delta A$ positive and reaching up to +0.5 Gyr), although the effect is small (see also fig.13 of DM19). 

However, it is conceivable that a much more significant improvement is achieved in some regions of the parameter space. For instance, the age derived for the most metal-poor LEGACY star, \object{KIC 8760414}, is revised upwards by almost 4 Gyrs using the relations making use of [Fe/H], and is no longer severely underestimated compared to the seismic result. DM19 cautioned that their relations are not robust for [Fe/H] $\lesssim$ --0.8. Whether the improvement observed for \object{KIC 8760414} is only a fluke or indicates that the relations are still of some applicability for metal-poor stars must be investigated further. The improvement arising from the use of 2D or 3D relations might be buried in the noise for the other stars in our sample with parameters closer to solar.

\subsection{Age inferences from abundance data for an extended sample}\label{sect_discussion_age_extended_sample}

To further examine the performance of the abundance-age relations proposed by DM19, we consider an extended sample of {\it Kepler} dwarfs and subgiants with a homogeneous determination in the literature of both the abundances and the seismic properties. Namely, we select stars in common between \citet{brewer16} and \citet{serenelli17}. Abundances of the key elements Sr and Y are not provided by \citet{bruntt12}. Note that \citet{brewer16} empirically corrected their abundance dataset for trends as a function of $T_\mathrm{eff}$. We use in the following the seismic results of \citet{serenelli17} based on SDSS data. We ignore stars with $T_\mathrm{eff}$ taken from \citet{brewer16} below 5300 K because they fall outside the validity domain of the calibrations of DM19. Four stars exhibit anomalously high abundance ratios involving yttrium: we exclude \object{KIC 9025370} because it is an SB2 (N17) and \object{KIC 11026764} because significantly lower values were obtained by \citet{metcalfe10}. We retain the remaining two stars: \object{KIC 3733735} and \object{KIC 9812850}. No additional information about a general enhancement of the $s-$process elements\footnote{For the very young ($\sim$0.5 Gyr) star \object{KIC 3733735}, only [Y/Al] is very high, which rather points to a problem with the Al abundance.} is available, but, in any case, the ages inferred from the high Y abundances are all negative and therefore not considered further. We finally end up with a total of 63 stars. It should be noted that they are on average younger ($\sim$3 Gyrs) and more evolved ($\log g$ values down to 3.5) than our LEGACY sample. The former feature likely explains the large proportion ($\sim$20\%) of unphysical (negative) ages obtained. Following \citet{battistini16}, we simply flag stars older than 8 Gyrs as thick-disc members. There are only four stars fulfilling this criterion.

\begin{figure}[h!]
\centering
\includegraphics[trim=110 200 185 130,clip,width=0.7\hsize]{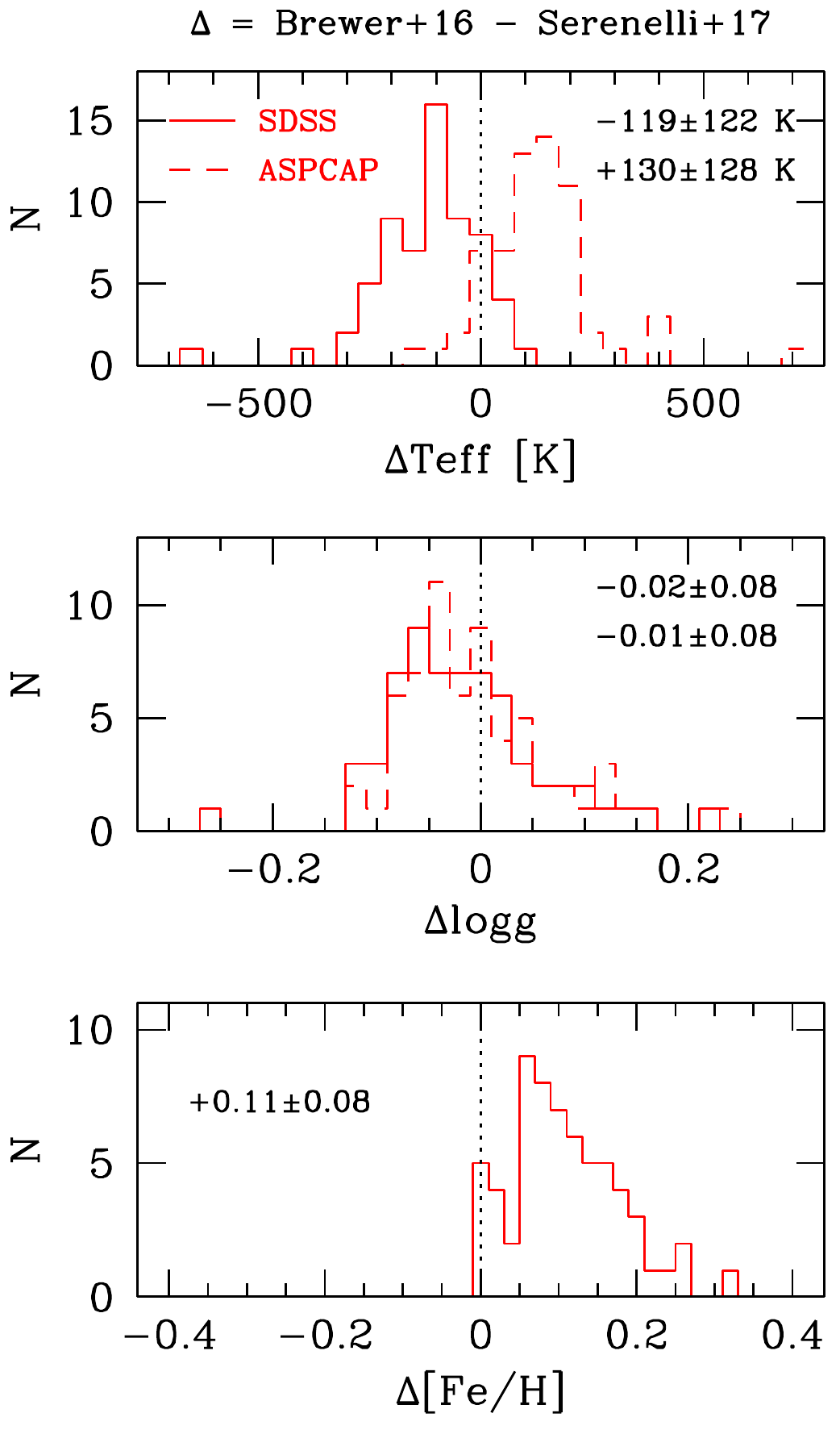}
\caption{Differences between the stellar parameters of \citet{brewer16} and \citet{serenelli17}. The mean values for the SDSS and ASPCAP $T_\mathrm{eff}$ scales are given in each panel. We assumed the ASPCAP [M/H] values that are the overall scaled-solar abundances.}
\label{fig_comparison_B16_APOKASC}
\end{figure}

Three limitations must be kept in mind. First, only seven abundance ratios can be computed from the data of \citet{brewer16} because Zn and Sr are not available. Second, the stellar parameters of \citet{serenelli17} are based on global seismic quantities and are expected to be a factor 2-3 less precise than those based on the modelling of the full set of frequencies or their separation ratios. Finally, we have argued in Sect.~\ref{sect_validation_parameters_literature} that for all but one LEGACY stars our spectroscopic parameters are consistent with those adopted by C17 and SA17 for their seismic modelling. For the remaining star, \object{KIC 9965715}, we performed a new seismic analysis using our $T_\mathrm{eff}$ and [Fe/H] values (Appendix \ref{sect_seismic_analysis_KIC9965715}). In contrast, here there are systematic differences between the classical parameters adopted for the determinations of the chemical abundances and ages (see Fig.~\ref{fig_comparison_B16_APOKASC}). For the stars selected, we find that the $T_\mathrm{eff}$ scale of \citet{brewer16} is on average $\sim$120 K cooler than the SDSS one. For completeness, we also show the comparison with the ASPCAP $T_\mathrm{eff}$ scale. It supports our previous conclusion that these $T_\mathrm{eff}$ values are too low. In addition, the metallicities of \citet{brewer16} are $\sim$0.11 dex larger, which is also fully consistent with our finding that the ASPCAP values may be underestimated by this amount (Sect.~\ref{sect_validation_parameters_literature}). In contrast, there is a satisfactory agreement between the spectroscopic and seismic gravities, as indeed anticipated \citep[see][]{brewer15}.

\begin{figure}[h!]
\centering
\includegraphics[trim=150 180 260 130,clip,width=0.7\hsize]{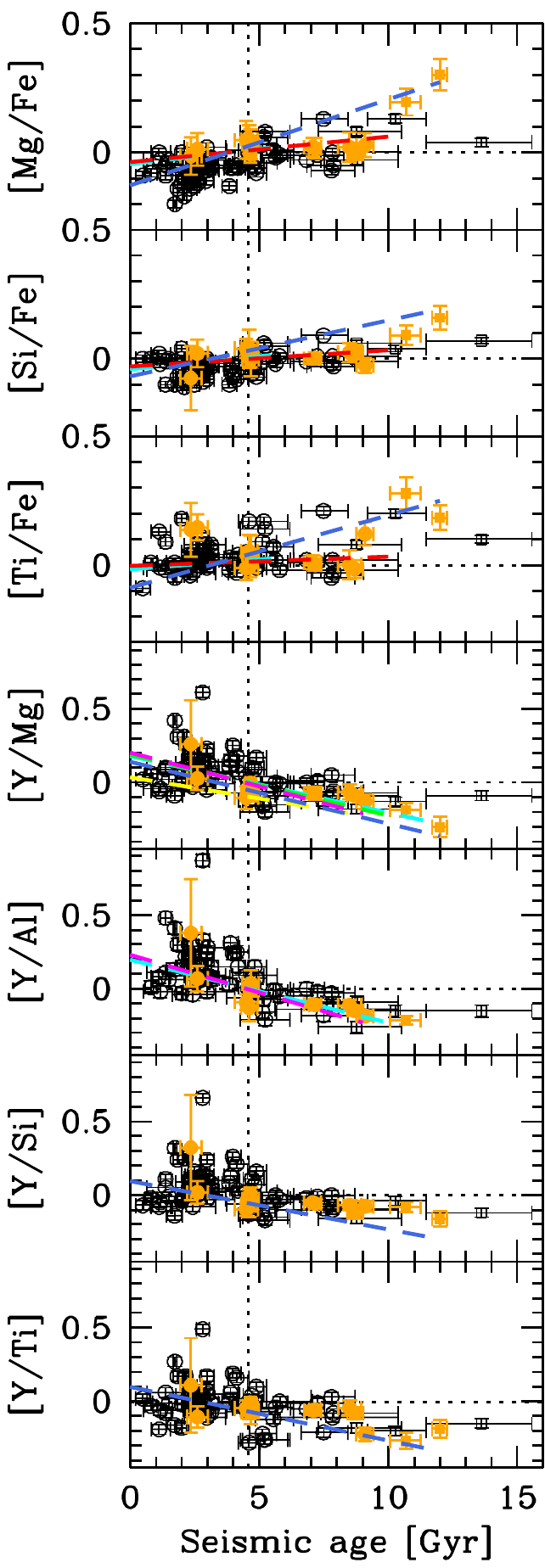}
\caption{Abundance-age trends for our data ({\it filled symbols}) and those of \citet[][{\it open symbols}]{brewer16}. Note that the ages were derived by different means: detailed modelling of the oscillation frequencies and from global seismic quantities, respectively. The very high [Y/Al] abundance ratio ($\sim$+1.35) obtained by \citet{brewer16} for \object{KIC 3733735} is off scale. The thin- and thick-disc stars are shown as circles and squares, respectively. Linear 1D relations in the literature are overplotted as dashed lines: \citet[][{\it red}]{bedell18}, DM19 ({\it blue}), \citet{nissen16} and \citet[][{\it cyan}]{nissen20}, \citet[][{\it magenta}]{spina18}, \citet[][{\it yellow}]{titarenko19}, and \citet[][{\it green}]{tucci_maia16}. The solar age is shown as a vertical dotted line. }
\label{fig_age_indicators_extended_sample}
\end{figure}

\begin{figure*}[h!]
\centering
\includegraphics[trim=25 170 20 160,clip,width=1.0\hsize]{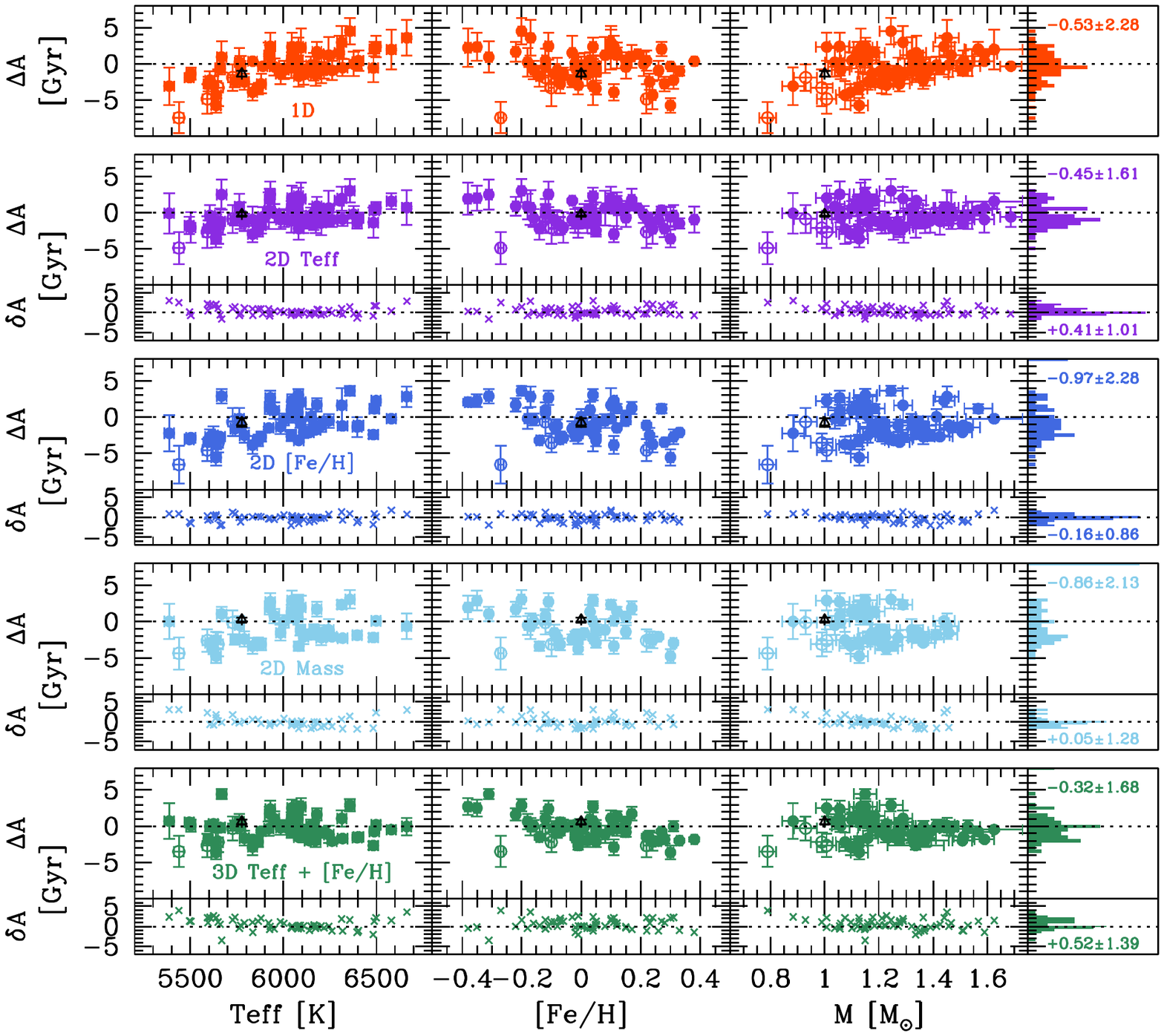}
\caption{Same as Fig.~\ref{fig_comparison_DM19_mean_ages_vs_parameters}, but for the extended sample.}
\label{fig_comparison_DM19_extended_mean_ages_common_vs_parameters}
\end{figure*}

Figure \ref{fig_age_indicators_extended_sample} shows the abundance-age trends for our data and the subsample of 63 stars from \citet{brewer16}. As shown in Fig.~\ref{fig_comparison_DM19_extended_mean_ages_common_vs_parameters}, the analysis of this much larger sample broadly confirms the conclusions drawn above for the LEGACY stars. Namely, the scatter of the deviations between the abundance- and seismic-based ages is of the order of 2 Gyrs and an improvement of up to $\sim$0.5 Gyr occurs with 2D or 3D relations. Noteworthy is the fact that the 2D $T_\mathrm{eff}$ and 3D relationships (the latter also involving $T_\mathrm{eff}$) quite efficiently erase the clear trend between the age difference and $T_\mathrm{eff}$. The lack of improvement when incorporating [Fe/H] as an additional variable might be due to the near-solar metallicity range. We find that negative $\delta A$ values for the 2D [Fe/H] calibration are generally associated to the most evolved stars ($\log g$ $\lesssim$ 4.1). One can notice in the panels as a function of mass in Fig.~\ref{fig_comparison_DM19_extended_mean_ages_common_vs_parameters} two sequences running almost parallel to the $x$-axis at $\Delta A$ $\sim$ --2 and +2 Gyrs (best seen for the 2D Mass relation). If real, there is no clear explanation for this behaviour. Once again, the seismic ages appear overall systematically larger by $\sim$0.6 Gyr. 

\section{Summary and conclusions}\label{sect_conclusions}

\subsection{Implication for further seismic modelling}\label{sect_conclusions_seismic_modelling}

 Our non-seismic parameters (especially $T_\mathrm{eff}$, [Fe/H], and [$\alpha$/Fe]) complement the superb {\it Kepler} oscillation data obtained for these bright stars and will aid further seismic modelling. These are important observational constraints that are necessary to exploit the full potential of asteroseismology \citep[see, e.g.][]{chaplin14,serenelli17,creevey12,valle18}. We classify two of our targets with an age exceeding 8 Gyrs as thick-disc members mostly based on chemodynamical arguments. Their chemical pattern significantly different from the solar mixture (in particular an enhancement of the $\alpha$ elements) must be taken into account to achieve sensible seismic inferences \citep[e.g.][]{ge15,li20}. Our results can also be useful in other more specific contexts, such as constraining the Galactic helium enrichment law \citep[e.g.][]{verma19} or improving the modelling of the outermost stellar layers \citep[e.g.][]{compton18}. Two Li-detected stars, \object{KIC 5184732} and \object{KIC 8694723}, also have a determination of their rotational period from {\it Kepler} light curves \citep{garcia14,karoff13}, which paves the way for a detailed theoretical modelling of their interior. 

We find evidence that a few of our targets are associated in a binary system with a large mass ratio. This suggests the existence in the {\it Kepler} LEGACY sample of a sizeable fraction of binaries. This is supported by the fact that N17 detected two SB2's in their sample of ten stars from snapshot observations only. It would not be surprising considering the ubiquity of companions in solar-like stars \citep[e.g.][]{duquennoy91}. This aspect certainly deserves further investigation (i.e. a dedicated RV monitoring).

As guidance for further ensemble seismic modelling of {\it Kepler} main-sequence stars, our comparison with the results of (semi-)automatic methods suggests that the study of \citet{buchhave15} is a valuable source of effective temperatures and metallicities. For the elemental abundances, the works of \citet{bruntt12} and \citet{brewer16} appear to be of similar quality for most elements. However, our results tend to be in closer agreement with those of the latter study. We confirm that the ASPCAP effective temperatures used for the APOKASC catalogue are systematically too cool for dwarfs. This is not completely unexpected given that the pipeline is naturally optimised for red giants \citep{garcia_perez16}. The consequence is that a grid-based modelling that makes use of the SDSS photometric $T_\mathrm{eff}$ scale may be more reliable for relatively unevolved {\it Kepler} targets, as already claimed by \citet[][]{serenelli17}. However, one can note in Fig.~\ref{fig_comparison_B16_APOKASC} the existence of a systematic discrepancy between the SDSS $T_\mathrm{eff}$ values and those of \citet{brewer16}. We find evidence in our sample that the APOKASC metallicities for dwarfs (solely based on ASPCAP) are underestimated by typically $\sim$0.1 dex. A similar conclusion is reached when considering the data of \citet{brewer16}. Correcting for such a bias would lead to a fractional improvement of a few percent in the stellar properties quoted in the APOKASC catalogue \citep[e.g. $\sim$5\% in age;][]{serenelli17}. As illustrated by our new seismic analysis of \object{KIC 9965715}, even larger changes can be expected if the quality of the classical parameters is mediocre.

\subsection{Stellar ages from chemical clocks beyond solar analogues}\label{sect_conclusions_abundance_indicators}

We find that the quality of stellar ages empirically determined from abundance data is currently limited by the precision of the calibrations, which is typically 1.5-2 Gyrs when venturing away from stars closely resembling our Sun. Constructing more complex relationships that take into account a dependency with one or two stellar parameters does improve the situation, if only slightly. For instance, incorporating $T_\mathrm{eff}$ as an additional variable globally improves the agreement between the abundance- and seismic-based ages for the APOKASC stars studied by \citet{brewer16}. The calibrations presently available appear significantly less precise than what can be achieved for stars akin to the Sun (precision often below 1 Gyr), but some abundance ratios may hopefully prove in the future more suitable than others outside the rather restricted domain defined by solar analogues. A concern is that applying relationships constructed from a specific abundance dataset (such as that of \citealt{delgado_mena17} used by DM19) is prone to the presence of study-to-study zero-point abundance offsets that are virtually unavoidable \citep[e.g.][]{jofre17}. More generally, the possible existence of systematic differences between isochrone and seismic ages must be investigated further to improve the accuracy of the relationships (see \citealt{berger20} for an example of comparison).

 Constructing empirical abundance-age calibrations allowing one to accurately infer the age of stars spanning a wide range in spectral type, evolutionary status and metallicity --- let alone born in different parts of the Galactic discs characterised by different star-formation histories --- appears to be a daunting endeavour. A large number of stars with solar-like oscillations will be detected by the TESS satellite, but this sample will be largely dominated by evolved subgiants \citep[][]{campante16}. The full accomplishment of this goal might thus await the release of seismic ages for thousands of bright main-sequence stars by the PLATO space mission \citep{rauer18} complemented by a high-quality determination of their surface chemical composition.

\begin{acknowledgements}
We would like to thank the anonymous referee for her/his careful reading of the manuscript and thoughtful suggestions. TM acknowledges financial support from Belspo for contract PRODEX PLATO mission development, and wishes to thank Werner Verschueren (Belspo) for allowing the funding of the mission at OHP. JM, AM, and EW acknowledge support from the ERC Consolidator Grant funding scheme (project ASTEROCHRONOMETRY, https://www.asterochronometry.eu, G.A. n. 772293). This work has made use of data from the European Space Agency (ESA) mission {\it Gaia} ({\tt https://www.cosmos.esa.int/gaia}), processed by the {\it Gaia} Data Processing and Analysis Consortium (DPAC, {\tt https://www.cosmos.esa.int/web/gaia/dpac/consortium}). Funding for the DPAC has been provided by national institutions, in particular the institutions participating in the {\it Gaia} Multilateral Agreement. This article made use of AIMS, a software for fitting stellar pulsation data, developed in the context of the SPACEINN network, funded by the European Commission’s Seventh Framework Programme. This research has also made use of NASA's Astrophysics Data System Bibliographic Services and the SIMBAD database operated at CDS, Strasbourg (France).
\end{acknowledgements}

\bibliographystyle{aa} 
\bibliography{ms} 

\begin{appendix}

\newpage
  
\section{Seismic analysis of \object{KIC 9965715}}\label{sect_seismic_analysis_KIC9965715}

\subsection{First approach ({\tt ASTEC+ADIPLS})}\label{sect_seismic_analysis_KIC9965715_1}
  
We approached the new interpretation of the data from \object{KIC 9965715} in two steps. The first step consisted in using a dense grid of BASTI stellar models \citep{basti2004,basti2018} to match the $T_\mathrm{eff}$, $\langle \Delta \nu \rangle$, and $\nu_\mathrm{max}$. The global seismic quantities are taken from \citet{lund17}. These models use a chemical enrichment law, and we refer to the relevant papers for details of the physics. We use a simple Bayesian approach in 2D (mass, age) on grids of different metallicities with mild priors that have little influence on our results: mass is restricted to [0.8,1.3] M$_{\sun}$, age is restricted to [0,10] Gyrs, and we use a prior on the evolution state defined as the relative amount of time spent in the specific phase of the star's life (main sequence, subgiant, and giant). The uncertainties are obtained by marginalising over the mass and age, and are defined as the 68\% confidence interval centred on the median values. We also record the model observables and so we can define an optimal surface gravity $\log g$ = 4.280$\pm$0.004 where we have also accounted for the different metallicity value.   

In a second step, we use the fitted parameters from the first step as initial starting points for a detailed stellar modelling. The use of the global seismic quantities brings important constraints on certain stellar parameters such as the surface gravity or density. However, for the best precision and improved accuracy (scaling relations are still subject to some systematic errors) on other parameters such as mass and age, we must use the information contained in the detailed seismic data. We adopted a similar approach to that presented in C17 by using the frequency ratios ($r_{01}, r_{02}$) in the optimisation. These values are calculated directly from the individual frequencies from \citet{lund17}, and we derive the covariance matrix $C$ by performing simple Monte-Carlo-like simulations.

In the optimisation approach we calculate a Likelihood 
\begin{equation}
\mathcal{L} = e^{-\frac{\chi^2}{2}}
\label{eqn:likelihood}
\end{equation}
where
\begin{equation}
\chi^2 = (x - x_M)^T C^{-1} (x - x_M).
\label{eqn:chisq}
\end{equation}

Here $x$ are the observational constraints and $x_M$ are the corresponding observables from the model. We additionally included the constraint of the mean large and small frequency separations, $\langle \Delta \nu \rangle$ and $\langle \delta \nu \rangle$, by calculating this value within the range of the observed frequencies. We also included the metallicity and $T_\mathrm{eff}$ constraint.

For this second step of stellar modelling, we used the Aarhus STellar Evolution Code ({\tt ASTEC}) and the Adiabatic Pulsation code ({\tt ADIPLS}) from \citet{jcd2008a,jcd2008b}. The set-up of the physics is as follows: we assumed a non-rotating, non-magnetic star. The opacities and equation-of-state tables are taken from the OPAL collaboration \citep{ir1996,rn2002}. Nuclear reaction rates were taken from \citet{angulo1999} and we include the values obtained by the LUNA collaboration for the $^{14}$N(p,$\gamma$)$^{15}$O reaction \citep{formicola2004}. We use the solar mixture of \citet{gs98}. Apart from the fixed physics described above, the only other parameters that control the evolution of the star are the mass, $M$, and the initial metallicity and helium mass fraction, $Z_i$ and $Y_i$, where $X_i+Y_i+Z_i = 1$, and $X$ refers to the hydrogen mass fraction. We also need to set the mixing-length parameter $\alpha$, defined as the ratio of the mean-free path of a fluid element, and the pressure scale height \citep[see][]{bv1958}. The final parameter that controls the current structure of the model is the age.

\begin{figure}
  \centering
\includegraphics[trim=35 0 80 520,clip, width=0.5\textwidth]{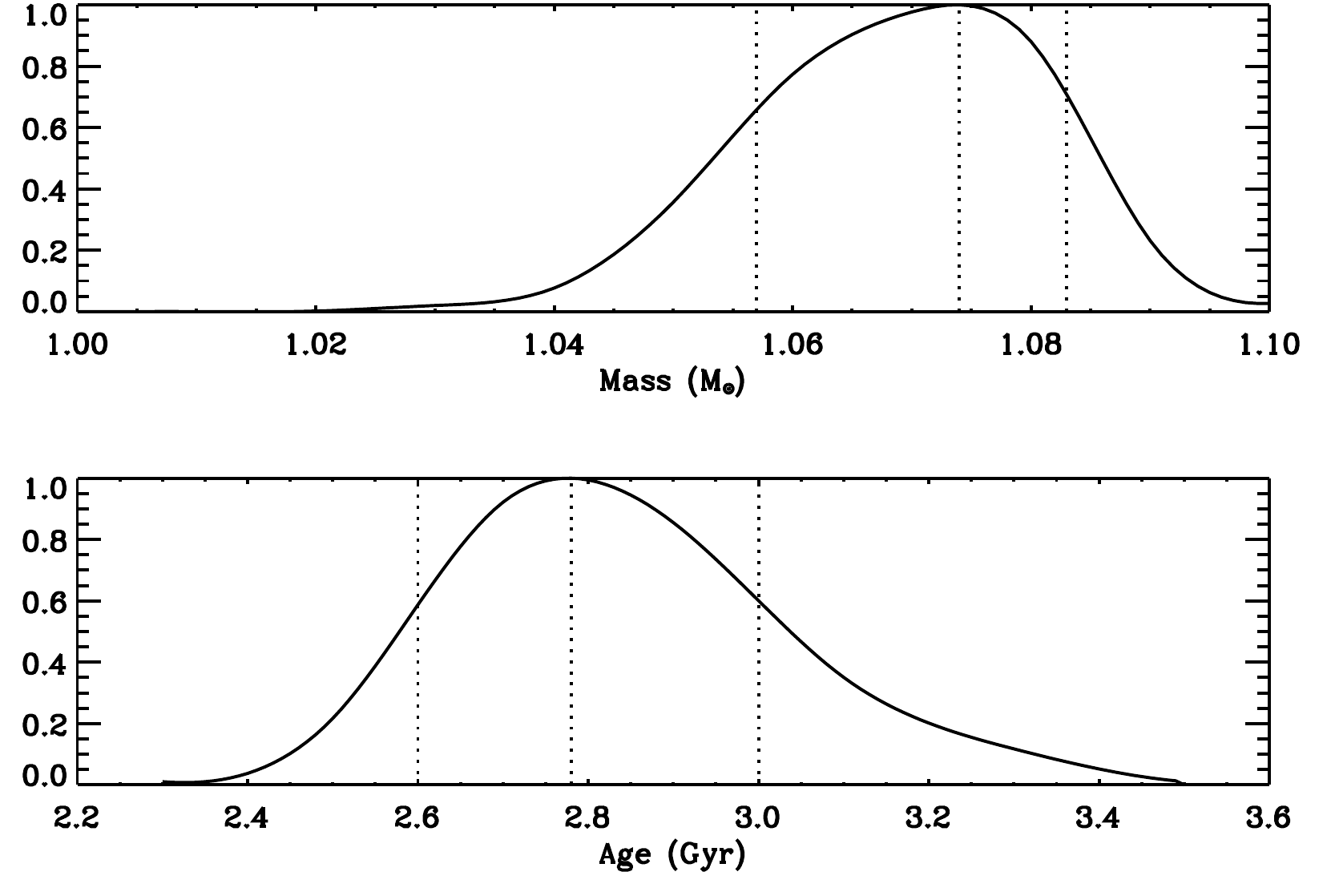}
\caption{Marginalised distributions of the mass and the age for the set ($Y_i,\alpha$) = (0.287,1.94), scaled so the maximum is equal to unity. The adopted stellar parameter (maximum likelihood) and the lower and upper values (68\% confidence interval) are shown by the dashed lines.\label{fig:orlagh}}
\end{figure}

We approached our optimisation of the parameters by building several small 3D grids ($M$, $Z_i$, and age) while fixing ($Y_i,\alpha$) at discrete values. We show some of these results in Table~\ref{tab:orlagh}, where the first set of ($Y_i,\alpha$) are equivalent to the values used in the BASTI library. The parameters are defined as the maximum likelihood ones, and the uncertainties  are calculated as the 68\% confidence interval with the highest probability, as shown by the dashed lines in Fig.~\ref{fig:orlagh}. We find that the mass varies between 1.05-1.13 M$_{\sun}$ with a strong dependency on $Y_{i}$. The optimal age remains in the range of between 2.7-3.3 Gyrs. We adopt the values from the third column and, to account for the other parameters, we add the differences to the uncertainties in quadrature, to obtain $M$ = 1.07$\pm$0.04 M$_{\sun}$ and an age of 2.8$\pm$0.4 Gyrs. 

\begin{table}
\begin{center}
  \caption{Stellar parameters for \object{KIC 9965715} using the first approach. The bottom rows give the adopted values. The solar global parameters are from IAU 2015 Resolution B3 \citep{mamajek15}.
    \label{tab:orlagh}}
\begin{tabular}{lc|ccc}
\hline\hline
($Y_i,\alpha$)      &               & (0.257, 2.14)      & (0.257, 2.04)       & (0.287, 1.94) \\
\hline             
$M$                 & [M$_{\sun}$]  & 1.12$\pm$0.01      & 1.11$\pm$0.01        & 1.07$_{-0.02}^{+0.01}$\\
$R$                 & [R$_{\sun}$]  & 1.29$\pm$0.01      & 1.29$\pm$0.01         & 1.26$\pm$0.01\\
Age                 & [Gyr]         & 3.1$^{+0.2}_{-0.1}$  & 3.20$_{-0.35}^{+0.05}$  & 2.8$\pm$0.2  \\
$\log\cal{L}$       &               & --10.1             & --9.2                & --8.3\\
\hline
$M$ & [M$_{\sun}$]   & \multicolumn{3}{c}{1.07$\pm$0.04}\\
Age                 & [Gyr]         & \multicolumn{3}{c}{2.8$\pm$0.4}\\
$\log g$            &              & \multicolumn{3}{c}{4.280$\pm$0.004}\\
\hline
\end{tabular}
\end{center}
\end{table}

\subsection{Second approach ({\tt CLES+LOSC})}\label{sect_seismic_analysis_KIC9965715_2}
    
The stellar parameters of \object{KIC 9965715} have also been estimated by using the open-source code {\tt AIMS} \citep[Asteroseismic Inference on a Massive Scale;][]{Reese2016,Lund_Reese2018,Rendle2019} that implements a Bayesian inference approach. {\tt AIMS} evaluates the posterior distributions of stellar parameters using a Markov Chain Monte Carlo (MCMC) ensemble sampler \citep{Foreman-Mackey2013}, and selects stellar models that best fit observational data by interpolating (evolutionary tracks and frequencies) in a pre-computed grid. 

The grid of stellar models at the base of the procedure was computed using the code {\tt CLES} \citep{cles,cles2}, and the oscillation frequencies of radial ($\ell=0$) and non-radial ($\ell=1,\,2$) modes for each stellar model using the adiabatic oscillation code {\tt LOSC} \citep{losc}. The stellar models computation follows the evolution from the pre-main sequence up to the point where the hydrogen content at the stellar centre, $X_\mathrm{c}$, is below 10$^{-6}$. The prescriptions for the input physics are the following: nuclear reaction rates of \citet{Adelberger_11} or from NACRE \citep{angulo1999} if not available. The equation of state is FreeEoS \citep{Irwin_2012}. We adopt the solar metal mixture from \cite{asplund09}, and opacity tables were built combining OPAL \citep{ir1996} opacity values at high temperature with those of Wichita State University for low temperature domain \citep{lowTOP2005}. Atmospheric boundary conditions are provided by the Krishna-Swamy's ${\rm T(\tau)}$ law \citep{KS66} and the corresponding atmosphere structure is added on the top of the interior model. Convection is treated with the ``mixing length'' formalism \citep{CG68}. The corresponding $\alpha$ parameter was kept fixed for all the grids, and was derived from the solar calibration ($\alpha$ = 2.338) with the same physics and including microscopic diffusion of chemical elements \citep{Thoul_94}. 
 
We assume extra-mixing at the boundaries of convective regions introducing core-overshooting ($ov$ = 0.1 and 0.2 $H_\mathrm{p}$ --- pressure scale height), and undershooting below the convective envelope ($\sim$0.2 $H_\mathrm{p}$). In both cases, the chemical mixing is instantaneous and the temperature gradient in the mixed region is the radiative one. Finally, microscopic diffusion of chemical elements has been included in one of the grids. 

The grids are parameterised by mass (from 0.9 to 1.8 M$_\sun$ with a step of 0.01 M$_\sun$) and initial $\mathrm{[Fe/H]}$ (from --0.45 to 0.00, with a step of 0.05 dex), assuming an enrichment law $\Delta\,Y/\Delta\, Z$ provided by the solar calibration and a primordial He abundance, $Y_\mathrm{p}$, of 0.2485 \citep{komatsu11}.  

The stellar parameters of \object{KIC 9965715} are derived using these grids and the following observational constraints: $T_\mathrm{eff}$ and $\mathrm{[Fe/H]}$, as classical constraints, as well as the average $\mathrm{\nu_\mathrm{max}}$ and the individual frequencies of 40 modes ($\ell=0,\,1,$ and 2) of \citet{lund17} as seismic data. We also perform a second set of optimisations using as seismic information the values of frequency separation ratios ($r_{01},\, r_{10},\, r_{02}$) and the frequency of lowest order radial mode. In the former case, we use the two-terms surface effect correction from \cite{Ball_Gizon2014}. 

The analysis of the results allows to rule out models with an extra-mixing as large as 0.2 $H_\mathrm{p}$. On the other hand, the overshooting below the convective envelope has no effect on the stellar parameters derived. From ``Bayesian Evidence'' values, models including microscopic diffusion of helium and metals are clearly preferred over those without it. 

The results of the different optimisations we perform indicate an initial chemical composition close to the solar one. So, stellar parameters based on non-diffusion grids tend to deviate from the observed surface chemical composition. A good match of surface composition can be obtained at the cost of a mediocre fit of the seismic properties. The stellar parameters based on frequency separation ratios are in good agreement with those derived using the individual frequencies. The parameters for each optimisation are defined from the posterior distributions, as the maximum likelihood ones, and the uncertainties are calculated as the 68\% confidence interval with the highest probability. We summarise these results in Table \ref{tab:josefina} in two sets of stellar parameters depending on whether they are based on the model grids with or without microscopic diffusion. We adopt the values of the first column and inflate the uncertainties to take into account that the target could be up to 1 Gyr older if the constraint on the surface chemical composition is relaxed.

\begin{table}
\begin{center}
  \caption{Stellar Parameters for \object{KIC 9965715} using the second approach. The bottom rows give the adopted values. The solar global parameters are from IAU 2015 Resolution B3 \citep{mamajek15}.
      \label{tab:josefina}}
\begin{tabular}{ll|cc}
\hline\hline
Grids             &  & With diffusion &  Without diffusion \\
\hline
$M$      & [M$_{\sun}$] &  1.20$\pm$0.01     & 1.166$\pm$0.002 \\
Age      & [Gyr]       &  2.41$\pm$0.14     & 3.37$\pm$0.06  \\
$R$      & [R$_{\sun}$] &  1.315$\pm$0.005   & 1.310 $\pm$0.004\\
$\log g$ &             &  4.279$\pm$0.001   & 4.270$\pm$0.001\\
${Z_0}$  &             &  0.0119$\pm$0.0007 & 0.0119$\pm$0.0003\\
${Y_0}$  &             &  0.2632$\pm$0.0009 &  0.2632$\pm$0.0003\\ 
\hline
$M$ & [M$_{\sun}$] & \multicolumn{2}{c}{1.20$^{+0.01}_{-0.13}$}\\
Age  & [Gyr]    & \multicolumn{2}{c}{2.41$^{+1.0}_{-0.14}$}\\
$\log g$      &   & \multicolumn{2}{c}{4.279$^{+0.001}_{-0.01}$}\\
\hline\end{tabular}
\end{center}
\end{table}

\section{Abundance results}\label{sect_abundance results}

\begin{table*}[h!]
\scriptsize
  \caption{Abundance results for the stars in our sample.}
\centering
\begin{tabular}{l|ccccccc}
\hline\hline
                                       & \multicolumn{1}{c}{\object{KIC 3656476}} & \multicolumn{1}{c}{\object{KIC 5184732}} & \multicolumn{1}{c}{\object{KIC 6106415}} & \multicolumn{1}{c}{\object{KIC 6603624}} & \multicolumn{1}{c}{\object{KIC 7871531}} & \multicolumn{1}{c}{\object{KIC 7970740}} & \multicolumn{1}{c}{\object{KIC 8006161}} \\
\hline
$[$Fe/H$]$\tablefootmark{a}            &     0.27$\pm$0.03 (33+6)                 &   0.41$\pm$0.04 (33+6)                   & --0.05$\pm$0.03 (39+5)                   &   0.27$\pm$0.04 (33+6)                   & --0.19$\pm$0.03 (34+5)                   & --0.47$\pm$0.03 (34+6)                   &   0.35$\pm$0.08 (28+5)               \\     
$\log \epsilon_\sun$(Li)                & $<$0.9 (1)                   &   2.27$\pm$0.07 (1)                &   2.49$\pm$0.06 (1)                & $<$1.0 (1)                         & $<$0.6 (1)                         & $<$0.6 (1)                        & $<$0.9 (1)    \\     
$[$\ion{C}{i}/Fe$]$                    &   --0.02$\pm$0.04 (2)                    & --0.10$\pm$0.06 (2)                      & --0.05$\pm$0.04 (2)                      &   0.01$\pm$0.05 (2)                      &   0.00$\pm$0.06 (1)                      &   0.04$\pm$0.06 (1)                      & --0.04$\pm$0.09 (1)                \\       
$[$\ion{Na}{i}/Fe$]$                   &     0.04$\pm$0.03 (2)                    &   0.16$\pm$0.04 (2)                      & --0.03$\pm$0.03 (2)                      &   0.09$\pm$0.04 (2)                      &   0.02$\pm$0.05 (2)                      &   0.03$\pm$0.05 (2)                      &   0.25$\pm$0.06 (2)   \\                    
$[$\ion{Mg}{i}/Fe$]$                   &   --0.01$\pm$0.03 (2)                    &   0.05$\pm$0.08 (2)                      & --0.01$\pm$0.04 (2)                      &   0.02$\pm$0.06 (2)                      &   0.03$\pm$0.05 (2)                      &   0.20$\pm$0.06 (2)                      &   0.05$\pm$0.06 (2)      \\                 
$[$\ion{Al}{i}/Fe$]$                   &     0.05$\pm$0.03 (2)                    &   0.03$\pm$0.05 (2)                      &   0.00$\pm$0.03 (2)                      &   0.07$\pm$0.04 (2)                      &   0.08$\pm$0.03 (2)                      &   0.23$\pm$0.04 (2)                      &   0.11$\pm$0.07 (2)    \\                   
$[$\ion{Si}{i}/Fe$]$                   &     0.03$\pm$0.03 (6)                    &   0.04$\pm$0.04 (5)                      &   0.00$\pm$0.04 (6)                      &   0.02$\pm$0.04 (5)                      & --0.03$\pm$0.03 (6)                      &   0.09$\pm$0.04 (6)                      &   0.04$\pm$0.08 (5)   \\                    
$[$\ion{Ca}{i}/Fe$]$                   &   --0.05$\pm$0.03 (3)                    & --0.02$\pm$0.04 (3)                      &   0.00$\pm$0.04 (4)                      & --0.02$\pm$0.04 (2)                      &   0.10$\pm$0.05 (3)                      &   0.18$\pm$0.04 (2)                      & --0.03$\pm$0.05 (2)   \\                    
$[$\ion{Sc}{ii}/Fe$]$\tablefootmark{b} &     0.02$\pm$0.03 (3)                    &   0.01$\pm$0.04 (3)                      & --0.01$\pm$0.04 (3)                      &   0.02$\pm$0.05 (3)                      &   0.02$\pm$0.04 (3)                      &   0.18$\pm$0.03 (3)                      &   0.10$\pm$0.10 (3)   \\                    
$[$\ion{Ti}{i}/Fe$]$                   &   --0.01$\pm$0.04 (6)                    & --0.01$\pm$0.05 (6)                      &   0.00$\pm$0.05 (6)                      &   0.00$\pm$0.06 (6)                      &   0.12$\pm$0.05 (7)                      &   0.28$\pm$0.07 (6)                      &   0.05$\pm$0.07 (5)  \\                     
$[$\ion{V}{i}/Fe$]$\tablefootmark{b}   &     0.02$\pm$0.04 (4)                    &   0.03$\pm$0.05 (4)                      & --0.01$\pm$0.04 (2)                      &   0.01$\pm$0.05 (4)                      &   0.16$\pm$0.07 (4)                      &   0.19$\pm$0.06 (4)                      &   0.14$\pm$0.07 (4)   \\                    
$[$\ion{Cr}{i}/Fe$]$                   &   --0.02$\pm$0.03 (2)                    &   0.02$\pm$0.04 (2)                      & --0.04$\pm$0.05 (3)                      & --0.02$\pm$0.05 (2)                      &   0.01$\pm$0.03 (3)                      &   0.02$\pm$0.05 (3)                      &   0.03$\pm$0.11 (2)  \\                     
$[$\ion{Mn}{i}/Fe$]$\tablefootmark{b}  &   --0.05$\pm$0.04 (3)                    &   0.03$\pm$0.06 (3)                      & --0.08$\pm$0.04 (3)                      & --0.01$\pm$0.06 (3)                      & --0.06$\pm$0.04 (3)                      & --0.09$\pm$0.06 (3)                      &   0.07$\pm$0.09 (3)  \\                     
$[$\ion{Co}{i}/Fe$]$\tablefootmark{b}  &   --0.04$\pm$0.04 (8)                    &   0.02$\pm$0.05 (6)                      & --0.06$\pm$0.05 (4)                      &   0.00$\pm$0.05 (5)                      &   0.00$\pm$0.04 (7)                      &   0.08$\pm$0.06 (9)                      &   0.09$\pm$0.07 (5)   \\                    
$[$\ion{Ni}{i}/Fe$]$                   &     0.02$\pm$0.03 (8)                    &   0.05$\pm$0.04 (6)                      & --0.02$\pm$0.04 (10)                     &   0.04$\pm$0.05 (10)                     & --0.04$\pm$0.04 (11)                     &   0.00$\pm$0.05 (11)                     &   0.10$\pm$0.06 (9)   \\                    
$[$\ion{Cu}{i}/Fe$]$\tablefootmark{b}  &     0.06$\pm$0.04 (2)                    &   0.11$\pm$0.07 (2)                      & --0.01$\pm$0.05 (2)                      &   0.10$\pm$0.07 (2)                      & --0.03$\pm$0.03 (2)                      &   0.04$\pm$0.03 (2)                      &   0.12$\pm$0.08 (1)   \\                    
$[$\ion{Zn}{i}/Fe$]$                   &     0.01$\pm$0.10 (3)                    &   0.03$\pm$0.13 (3)                      & --0.07$\pm$0.05 (3)                      &   0.08$\pm$0.09 (3)                      &   0.01$\pm$0.07 (3)                      &   0.14$\pm$0.09 (3)                      &   0.20$\pm$0.19 (3)   \\                    
$[$\ion{Sr}{i}/Fe$]$                   &   --0.14$\pm$0.06 (1)                    & --0.04$\pm$0.07 (1)                      & --0.03$\pm$0.06 (1)                      & --0.11$\pm$0.07 (1)                      & --0.09$\pm$0.06 (1)                      & --0.05$\pm$0.07 (1)                      &   0.00$\pm$0.17 (1)   \\                    
$[$\ion{Y}{ii}/Fe$]$                   &   --0.09$\pm$0.04 (4)                    & --0.06$\pm$0.05 (3)                      & --0.01$\pm$0.03 (4)                      & --0.04$\pm$0.04 (3)                      & --0.10$\pm$0.04 (2)                      &   0.01$\pm$0.03 (2)                      & --0.02$\pm$0.06 (3)  \\                     
$[$\ion{Ba}{ii}/Fe$]$                  &   --0.03$\pm$0.06 (1)                    & --0.08$\pm$0.07 (1)                      &   0.06$\pm$0.07 (1)                      & --0.03$\pm$0.07 (1)                      &   0.04$\pm$0.06 (1)                      &   0.04$\pm$0.06 (1)                      & --0.09$\pm$0.15 (1)   \\                    
$[$\ion{Ce}{ii}/Fe$]$                  &     0.00$\pm$0.03 (2)                    & --0.12$\pm$0.04 (2)                      &   0.08$\pm$0.06 (2)                      & --0.01$\pm$0.05 (2)                      &   0.09$\pm$0.08 (2)                      &   0.24$\pm$0.13 (2)                      &   0.04$\pm$0.08 (2)  \\                     
\hline                                     																																					    
$[$\ion{Sr}{i}/\ion{Mg}{i}$]$          &   --0.13$\pm$0.06                        & --0.09$\pm$0.08                          & --0.02$\pm$0.06                          & --0.13$\pm$0.07                          & --0.12$\pm$0.07                          & --0.24$\pm$0.07                          & --0.06$\pm$0.15   \\                        
$[$\ion{Sr}{i}/\ion{Al}{i}$]$          &   --0.19$\pm$0.06                        & --0.08$\pm$0.07                          & --0.03$\pm$0.06                          & --0.17$\pm$0.07                          & --0.17$\pm$0.06                          & --0.27$\pm$0.06                          & --0.12$\pm$0.21  \\                         
$[$\ion{Sr}{i}/\ion{Si}{i}$]$          &   --0.17$\pm$0.06                        & --0.08$\pm$0.07                          & --0.03$\pm$0.06                          & --0.13$\pm$0.07                          & --0.06$\pm$0.06                          & --0.14$\pm$0.07                          & --0.04$\pm$0.22   \\                        
$[$\ion{Sr}{i}/\ion{Ti}{i}$]$          &   --0.12$\pm$0.06                        & --0.03$\pm$0.06                          & --0.03$\pm$0.06                          & --0.11$\pm$0.07                          & --0.21$\pm$0.07                          & --0.32$\pm$0.08                          & --0.06$\pm$0.15  \\                         
$[$\ion{Sr}{i}/\ion{Zn}{i}$]$          &   --0.15$\pm$0.12                        & --0.07$\pm$0.14                          &   0.04$\pm$0.07                          & --0.18$\pm$0.11                          & --0.10$\pm$0.09                          & --0.18$\pm$0.10                          & --0.21$\pm$0.22   \\                        
$[$\ion{Y}{ii}/\ion{Mg}{i}$]$          &   --0.09$\pm$0.04                        & --0.10$\pm$0.08                          &   0.00$\pm$0.03                          & --0.06$\pm$0.06                          & --0.13$\pm$0.06                          & --0.19$\pm$0.05                          & --0.07$\pm$0.05   \\                          
$[$\ion{Y}{ii}/\ion{Al}{i}$]$          &   --0.15$\pm$0.04                        & --0.09$\pm$0.06                          & --0.01$\pm$0.03                          & --0.11$\pm$0.04                          & --0.18$\pm$0.04                          & --0.22$\pm$0.03                          & --0.13$\pm$0.09    \\                       
$[$\ion{Y}{ii}/\ion{Si}{i}$]$          &   --0.12$\pm$0.04                        & --0.10$\pm$0.05                          & --0.01$\pm$0.03                          & --0.07$\pm$0.04                          & --0.08$\pm$0.04                          & --0.08$\pm$0.04                          & --0.06$\pm$0.10   \\                        
$[$\ion{Y}{ii}/\ion{Ti}{i}$]$          &   --0.08$\pm$0.04                        & --0.05$\pm$0.06                          & --0.01$\pm$0.05                          & --0.05$\pm$0.06                          & --0.22$\pm$0.05                          & --0.27$\pm$0.06                          & --0.07$\pm$0.06  \\                         
$[$\ion{Y}{ii}/\ion{Zn}{i}$]$          &   --0.10$\pm$0.11                        & --0.09$\pm$0.13                          &   0.06$\pm$0.04                          & --0.12$\pm$0.09                          & --0.12$\pm$0.07                          & --0.13$\pm$0.09                          & --0.22$\pm$0.18   \\                        
\hline
\end{tabular}
\end{table*}

\addtocounter{table}{-1}
\addtocounter{subtable}{1}
\begin{table*}[h!]
\scriptsize
  \caption{Continued.}
\label{tab_abundance_results} 
\centering
\begin{tabular}{l|cccccc}
\hline\hline
                                        & \multicolumn{1}{c}{\object{KIC 8694723}} & \multicolumn{1}{c}{\object{KIC 8760414}} & \multicolumn{1}{c}{\object{KIC 9965715}} & \multicolumn{1}{c}{\object{KIC 12069424}} & \multicolumn{1}{c}{\object{KIC 12069449}} & \multicolumn{1}{c}{\object{KIC 12317678}}\\
                                        &                                          &                                          &                                          & \multicolumn{1}{c}{(\object{16 Cyg A}) }                       & \multicolumn{1}{c}{(\object{16 Cyg B})}  & \\
                                       
\hline
$[$Fe/H$]$\tablefootmark{a}             & --0.41$\pm$0.06 (22+4)                   & --0.95$\pm$0.05 (14+4)                   & --0.29$\pm$0.04 (20+4)                   &   0.11$\pm$0.02 (35+5)                    &    0.08$\pm$0.02 (35+5)                   & --0.19$\pm$0.12 (19+3)\\     
$\log \epsilon_\sun$(Li)           &   2.09$\pm$0.11 (1)                &   2.05$\pm$0.08 (1)                &   2.19$\pm$0.08 (1)                &   1.30$\pm$0.08 (1)                & $<$0.9 (1)                           & 2.98$\pm$0.15 (1) \\   
$[$\ion{C}{i}/Fe$]$                     &   0.01$\pm$0.09 (2)                      &   0.08$\pm$0.08 (1)                      &   0.03$\pm$0.06 (2)                      & --0.07$\pm$0.04 (2)                       &  --0.06$\pm$0.04 (2)                      & --0.08$\pm$0.18 (2) \\ 
$[$\ion{Na}{i}/Fe$]$                    & --0.01$\pm$0.06 (2)                      & --0.07$\pm$0.07 (1)                      &   0.04$\pm$0.04 (2)                      &   0.01$\pm$0.02 (2)                       &    0.01$\pm$0.02 (2)                      & --0.02$\pm$0.12 (2) \\ 
$[$\ion{Mg}{i}/Fe$]$                    &   0.03$\pm$0.06 (2)                      &   0.30$\pm$0.07 (2)                      &   0.01$\pm$0.07 (2)                      &   0.02$\pm$0.04 (3)                       &    0.02$\pm$0.04 (3)                      & --0.01$\pm$0.08 (2) \\ 
$[$\ion{Al}{i}/Fe$]$                    & --0.07$\pm$0.08 (1)                      & \multicolumn{1}{c}{...}                  & --0.03$\pm$0.07 (1)                      &   0.05$\pm$0.03 (2)                       &    0.05$\pm$0.02 (2)                      & --0.13$\pm$0.14 (1) \\ 
$[$\ion{Si}{i}/Fe$]$                    &   0.00$\pm$0.07 (6)                      &   0.16$\pm$0.05 (3)                      &   0.02$\pm$0.06 (6)                      &   0.00$\pm$0.02 (6)                       &    0.00$\pm$0.02 (6)                      & --0.07$\pm$0.13 (5) \\ 
$[$\ion{Ca}{i}/Fe$]$                    &   0.04$\pm$0.05 (3)                      &   0.21$\pm$0.06 (3)                      &   0.06$\pm$0.04 (3)                      & --0.01$\pm$0.04 (3)                       &  --0.01$\pm$0.03 (3)                      &   0.23$\pm$0.06 (2) \\ 
$[$\ion{Sc}{ii}/Fe$]$\tablefootmark{b}  &   0.04$\pm$0.06 (3)                      &   0.19$\pm$0.07 (1)                      &   0.03$\pm$0.05 (2)                      &   0.02$\pm$0.02 (3)                       &    0.04$\pm$0.02 (3)                      & --0.10$\pm$0.10 (2) \\ 
$[$\ion{Ti}{i}/Fe$]$                    &   0.04$\pm$0.08 (3)                      &   0.18$\pm$0.05 (2)                      &   0.15$\pm$0.05 (3)                      &   0.00$\pm$0.03 (5)                       &    0.01$\pm$0.03 (5)                      &   0.14$\pm$0.11 (1) \\ 
$[$\ion{V}{i}/Fe$]$\tablefootmark{b}    &   0.09$\pm$0.10 (2)                      & \multicolumn{1}{c}{...}                  &   0.25$\pm$0.17 (2)                      &   0.00$\pm$0.03 (5)                       &  --0.01$\pm$0.04 (5)                      & \multicolumn{1}{c}{...} \\ 
$[$\ion{Cr}{i}/Fe$]$                    & --0.10$\pm$0.08 (2)                      & --0.04$\pm$0.14 (2)                      &   0.07$\pm$0.04 (2)                      &   0.00$\pm$0.05 (3)                       &    0.01$\pm$0.05 (3)                      &   0.03$\pm$0.09 (2) \\ 
$[$\ion{Mn}{i}/Fe$]$\tablefootmark{b}   & --0.22$\pm$0.06 (2)                      & --0.32$\pm$0.05 (2)                      & --0.16$\pm$0.07 (2)                      & --0.04$\pm$0.03 (3)                       &  --0.03$\pm$0.03 (3)                      & --0.20$\pm$0.09 (1) \\ 
$[$\ion{Co}{i}/Fe$]$\tablefootmark{b}   & \multicolumn{1}{c}{...}                  & \multicolumn{1}{c}{...}                  & \multicolumn{1}{c}{...}                  & --0.04$\pm$0.04 (7)                       &  --0.01$\pm$0.03 (7)                      & --0.04$\pm$0.12 (1) \\ 
$[$\ion{Ni}{i}/Fe$]$                    & --0.06$\pm$0.06 (5)                      & --0.04$\pm$0.09 (6)                      & --0.09$\pm$0.05 (6)                      &   0.00$\pm$0.03 (10)                      &    0.00$\pm$0.02 (10)                     & --0.12$\pm$0.09 (4) \\ 
$[$\ion{Cu}{i}/Fe$]$\tablefootmark{b}   & --0.06$\pm$0.07 (1)                      & --0.17$\pm$0.07 (1)                      & --0.09$\pm$0.06 (1)                      &   0.03$\pm$0.02 (2)                       &    0.03$\pm$0.03 (2)                      & --0.24$\pm$0.11 (1) \\ 
$[$\ion{Zn}{i}/Fe$]$                    & --0.06$\pm$0.07 (3)                      &   0.01$\pm$0.12 (3)                      & --0.19$\pm$0.04 (2)                      &   0.03$\pm$0.05 (3)                       &    0.03$\pm$0.03 (3)                      & --0.12$\pm$0.15 (2) \\ 
$[$\ion{Sr}{i}/Fe$]$                    & --0.02$\pm$0.08 (1)                      &   0.05$\pm$0.07 (1)                      &   0.02$\pm$0.07 (1)                      & --0.10$\pm$0.06 (1)                       &  --0.10$\pm$0.06 (1)                      &   0.13$\pm$0.09 (1) \\ 
$[$\ion{Y}{ii}/Fe$]$                    & --0.03$\pm$0.07 (4)                      &   0.00$\pm$0.05 (3)                      &   0.04$\pm$0.07 (2)                      & --0.06$\pm$0.03 (4)                       &  --0.05$\pm$0.02 (4)                      &   0.25$\pm$0.28 (3) \\ 
$[$\ion{Ba}{ii}/Fe$]$                   &   0.20$\pm$0.18 (1)                      & --0.04$\pm$0.10 (1)                      &   0.16$\pm$0.11 (1)                      &   0.00$\pm$0.06 (1)                       &    0.00$\pm$0.06 (1)                      & \multicolumn{1}{c}{...} \\ 
$[$\ion{Ce}{ii}/Fe$]$                   &   0.06$\pm$0.07 (1)                      & \multicolumn{1}{c}{...}                  &   0.22$\pm$0.06 (1)                      & --0.01$\pm$0.03 (2)                       &    0.01$\pm$0.02 (2)                      &   0.20$\pm$0.08 (1) \\ 
\hline                                 
$[$\ion{Sr}{i}/\ion{Mg}{i}$]$           & --0.05$\pm$0.06                          & --0.25$\pm$0.08                          &   0.00$\pm$0.08                          & --0.11$\pm$0.07                           &  --0.12$\pm$0.06                          &   0.14$\pm$0.06    \\ 
$[$\ion{Sr}{i}/\ion{Al}{i}$]$           &   0.05$\pm$0.08                          & \multicolumn{1}{c}{...}                  &   0.05$\pm$0.08                          & --0.15$\pm$0.06                           &  --0.15$\pm$0.06                          &   0.26$\pm$0.12    \\ 
$[$\ion{Sr}{i}/\ion{Si}{i}$]$           & --0.01$\pm$0.08                          & --0.10$\pm$0.06                          &   0.00$\pm$0.07                          & --0.10$\pm$0.06                           &  --0.10$\pm$0.06                          &   0.20$\pm$0.12    \\ 
$[$\ion{Sr}{i}/\ion{Ti}{i}$]$           & --0.06$\pm$0.07                          & --0.13$\pm$0.06                          & --0.13$\pm$0.06                          & --0.10$\pm$0.06                           &  --0.11$\pm$0.06                          & --0.01$\pm$0.08    \\ 
$[$\ion{Sr}{i}/\ion{Zn}{i}$]$           &   0.04$\pm$0.09                          &   0.04$\pm$0.13                          &   0.20$\pm$0.07                          & --0.12$\pm$0.07                           &  --0.12$\pm$0.06                          &   0.25$\pm$0.17    \\ 
$[$\ion{Y}{ii}/\ion{Mg}{i}$]$           & --0.06$\pm$0.07                          & --0.30$\pm$0.07                          &   0.03$\pm$0.08                          & --0.07$\pm$0.05                           &  --0.07$\pm$0.04                          &   0.26$\pm$0.30    \\ 
$[$\ion{Y}{ii}/\ion{Al}{i}$]$           &   0.04$\pm$0.09                          & \multicolumn{1}{c}{...}                  &   0.07$\pm$0.09                          & --0.11$\pm$0.04                           &  --0.10$\pm$0.03                          &   0.38$\pm$0.37    \\ 
$[$\ion{Y}{ii}/\ion{Si}{i}$]$           & --0.03$\pm$0.09                          & --0.16$\pm$0.06                          &   0.02$\pm$0.08                          & --0.06$\pm$0.03                           &  --0.05$\pm$0.02                          &   0.32$\pm$0.36    \\ 
$[$\ion{Y}{ii}/\ion{Ti}{i}$]$           & --0.07$\pm$0.09                          & --0.19$\pm$0.06                          & --0.11$\pm$0.08                          & --0.06$\pm$0.04                           &  --0.06$\pm$0.04                          &   0.11$\pm$0.33    \\ 
$[$\ion{Y}{ii}/\ion{Zn}{i}$]$           &   0.03$\pm$0.08                          & --0.01$\pm$0.12                          &   0.23$\pm$0.06                          & --0.08$\pm$0.06                           &  --0.08$\pm$0.03                          &   0.36$\pm$0.20  \\ 
\hline
\end{tabular}
\tablefoot{The number in brackets gives the number of lines the abundance ratio is based on. \\
  \tablefoottext{a}{Weighted average of the \ion{Fe}{i}- and \ion{Fe}{ii}-based abundances.}
  \tablefoottext{b}{Corrected for HFS effects.}
}
\end{table*}

\newpage

  \section{Comparison with literature results for \object{16 Cyg AB}}\label{appendix_16_cyg_AB}

The detailed abundance pattern of the components in the well-studied binary system \object{16 Cyg AB} has recently been investigated with high precision by a number of studies. These are the two brightest solar analogues in the {\it Kepler} field. Noteworthy is the fact that the primary is slightly more metal-rich than the secondary. The following results were obtained as a function of increasing significance level: $\Delta$[Fe/H] (A--B) = +0.025$\pm$0.009 \citep{laws01}, +0.031$\pm$0.010 (N17), and +0.041$\pm$0.004 dex \citep{tucci_maia19}. The origin of this difference in metal content is unclear, but might be related to planetary formation and possibly subsequent engulfment. A Jupiter-mass planet is actually known to orbit \object{16 Cyg B} along a very eccentric ($e$ $\sim$ 0.6) orbit \citep{cochran97}.

 In addition to the analysis described in Sect.~\ref{sect_abundance_analysis}, we performed a differential analysis of \object{16 Cyg A} with respect to \object{16 Cyg B}. The parameters of \object{16 Cyg B} in Table \ref{tab_properties} were adopted. Our analysis procedures strictly follow those employed relative to solar for the other targets. Such an approach for binary components with closely similar parameters is expected to increase the precision of the results and, consequently, to better reveal subtle chemical differences \citep[e.g.][]{ramirez15}. We indeed find that it leads to a better precision with respect to the solar analysis. For instance, the line-to-line scatter in the iron abundances is divided by a factor of about two (here a mere $\sim$0.01 dex). The results are given in Table \ref{tab_results_16_Cyg_AB}. 

\begin{table}[h!]
\caption{Abundance results for \object{16 Cyg A} with respect to \object{16 Cyg B}.}
\label{tab_results_16_Cyg_AB} 
\centering
\begin{tabular}{l|crr}
\hline\hline
 &\multicolumn{3}{c}{\object{KIC 12069424} (16 Cyg A)}\\
\hline
$\Delta$$T_\mathrm{eff}$ [K]                    & \multicolumn{3}{c}{+50$\pm$11} \\
$\Delta$$\log g$ [cgs]\tablefootmark{a}        & \multicolumn{3}{c}{--0.065$\pm$0.005} \\
$\Delta$$\xi$ [km s$^{-1}$]                     & \multicolumn{3}{c}{+0.059$\pm$0.022} \\  
\hline
\multicolumn{4}{c}{}\\
      X                        & $N$   &  \multicolumn{1}{c}{$\Delta$$[$X/H$]$} & \multicolumn{1}{c}{$\Delta$$[$X/Fe$]$} \\
\hline
Fe\tablefootmark{b}           &  36+5 &   +0.022$\pm$0.010  &   \multicolumn{1}{c}{...}            \\
\ion{C}{i}                    &     2 &   +0.019$\pm$0.011  & --0.003$\pm$0.014\\
\ion{Na}{i}                   &     2 &   +0.033$\pm$0.013  &  +0.011$\pm$0.016\\
\ion{Mg}{i}                   &     3 &   +0.018$\pm$0.008  & --0.004$\pm$0.011\\
\ion{Al}{i}                   &     2 &   +0.027$\pm$0.016  &  +0.005$\pm$0.018\\
\ion{Si}{i}                   &     6 &   +0.030$\pm$0.009  &  +0.008$\pm$0.012\\
\ion{Ca}{i}                   &     3 &   +0.022$\pm$0.023  &  +0.000$\pm$0.024\\
\ion{Sc}{ii}\tablefootmark{c} &     3 &   +0.006$\pm$0.015  & --0.016$\pm$0.013\\
\ion{Ti}{i}                   &     5 &   +0.017$\pm$0.013  & --0.005$\pm$0.013\\
\ion{V}{i}\tablefootmark{c}   &     5 &   +0.032$\pm$0.030  &  +0.010$\pm$0.030\\
\ion{Cr}{i}                   &     3 &   +0.018$\pm$0.010  & --0.004$\pm$0.011\\
\ion{Mn}{i}\tablefootmark{c}  &     3 &   +0.015$\pm$0.016  & --0.007$\pm$0.016\\ 
\ion{Co}{i}\tablefootmark{c}  &     7 &   +0.003$\pm$0.030  & --0.019$\pm$0.030\\
\ion{Ni}{i}                   &    13 &   +0.023$\pm$0.012  &  +0.001$\pm$0.013\\
\ion{Cu}{i}\tablefootmark{c}  &     2 &   +0.025$\pm$0.015  &  +0.003$\pm$0.017\\
\ion{Zn}{i}                   &     3 &   +0.021$\pm$0.026  & --0.001$\pm$0.027\\
\ion{Sr}{i}                   &     1 &   +0.022$\pm$0.051  &  +0.000$\pm$0.051\\
\ion{Y}{ii}                   &     4 &   +0.019$\pm$0.026  & --0.003$\pm$0.025\\
\ion{Ba}{ii}                  &     1 &   +0.022$\pm$0.052  &  +0.000$\pm$0.051\\
\ion{Ce}{ii}                  &     2 &   +0.004$\pm$0.021  & --0.018$\pm$0.019\\
\hline
$\Delta$$[$\ion{Y}{ii}/\ion{Mg}{i}$]$    &  \multicolumn{3}{c}{ +0.001$\pm$0.026}\\
$\Delta$$[$\ion{Y}{ii}/\ion{Al}{i}$]$    &  \multicolumn{3}{c}{--0.008$\pm$0.030}\\
$\Delta$$[$\ion{Y}{ii}/\ion{Si}{i}$]$    &  \multicolumn{3}{c}{--0.011$\pm$0.026}\\
$\Delta$$[$\ion{Y}{ii}/\ion{Ti}{i}$]$    &  \multicolumn{3}{c}{ +0.002$\pm$0.026}\\
$\Delta$$[$\ion{Y}{ii}/\ion{Zn}{i}$]$    &  \multicolumn{3}{c}{--0.002$\pm$0.035}\\
$\Delta$$[$\ion{Sr}{i}/\ion{Mg}{i}$]$    &  \multicolumn{3}{c}{ +0.004$\pm$0.050}\\
$\Delta$$[$\ion{Sr}{i}/\ion{Al}{i}$]$    &  \multicolumn{3}{c}{--0.005$\pm$0.053}\\
$\Delta$$[$\ion{Sr}{i}/\ion{Si}{i}$]$    &  \multicolumn{3}{c}{--0.008$\pm$0.051}\\
$\Delta$$[$\ion{Sr}{i}/\ion{Ti}{i}$]$    &  \multicolumn{3}{c}{ +0.005$\pm$0.050}\\
$\Delta$$[$\ion{Sr}{i}/\ion{Zn}{i}$]$    &  \multicolumn{3}{c}{ +0.001$\pm$0.057}\\
\hline
\end{tabular}
\tablefoot{$N$ is the number of lines the abundance ratio is based on. \\
\tablefoottext{a}{Given by the seismic values (see Table \ref{tab_seismic_data}).}
\tablefoottext{b}{Weighted average of the \ion{Fe}{i}- and \ion{Fe}{ii}-based abundances.}
\tablefoottext{c}{Corrected for HFS effects.}
}
\end{table}

\begin{figure*}[h!]
\begin{minipage}[t]{0.5\textwidth}
\centering
\includegraphics[trim=135 175 100 135,clip,width=0.95\textwidth]{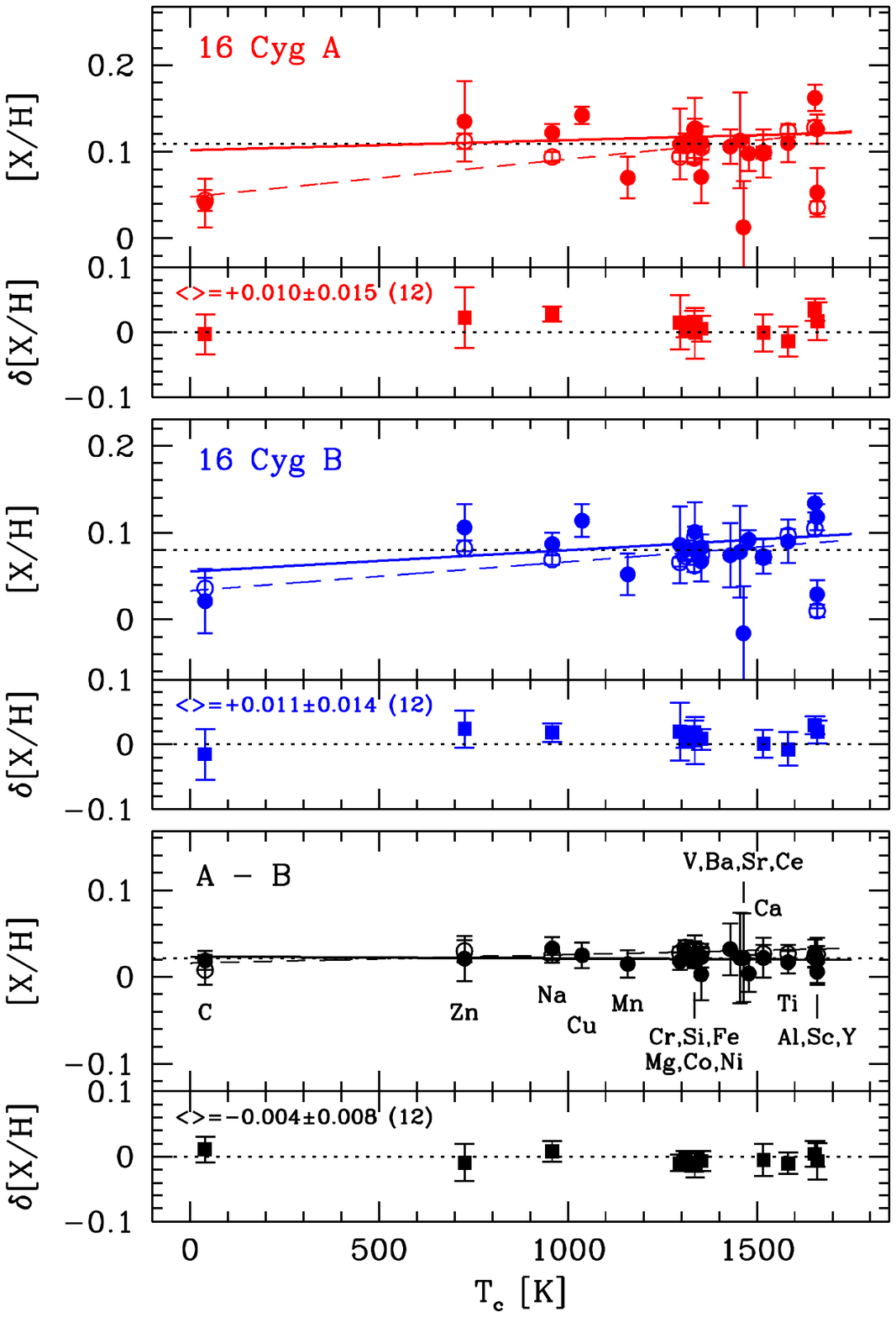}
\end{minipage}
\begin{minipage}[t]{0.5\textwidth}
\centering
\includegraphics[trim=135 175 100 135,clip,width=0.95\textwidth]{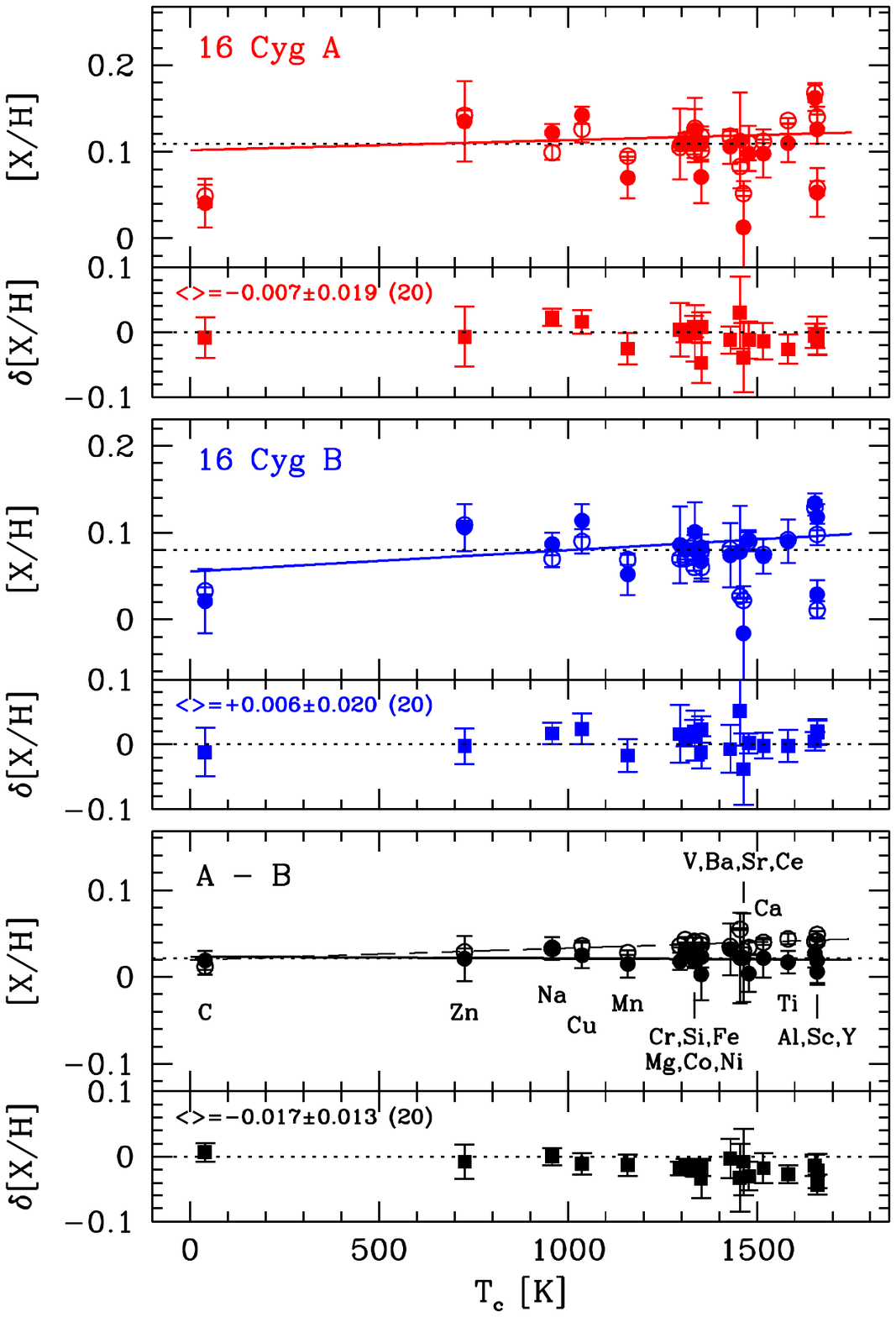}
\end{minipage}
\caption{Abundance patterns with respect to hydrogen, [X/H], and differences with respect to literature values (this study minus literature), $\delta$[X/H], as a function of $T_\mathrm{c}$. The left and right panels show the comparison with N17 and \citet{tucci_maia19}, respectively. We ignored the molecular-based C abundances of \citet{tucci_maia19}. The results for \object{16 Cyg A}, \object{16 Cyg B}, and \object{16 Cyg A} with respect to \object{16 Cyg B} are shown in the top, middle and bottom panels, respectively. Our [X/H] results and those in the literature are shown as filled and open circles, respectively. A dotted, horizontal line is drawn at our [Fe/H] value. The solid lines show the weighted, linear fit of [X/H] as a function of $T_\mathrm{c}$. The fits obtained in the literature for the full set of elements are overplotted as dashed lines (the trends for \object{16 Cyg AB} with respect to the Sun are not available in \citealt{tucci_maia19}). The average $\delta$[X/H] values are given in the relevant panels (the number of elements in common is indicated in brackets). To guide the eye, a dotted line is drawn at $\delta$[X/H] = 0.}
\label{fig_condensation_16_Cyg}
\end{figure*}

With respect to N17 and \citet{tucci_maia19}, we obtain a slightly cooler $T_\mathrm{eff}$ scale by $\sim$20 K on average. Figure \ref{fig_condensation_16_Cyg} shows a comparison between the abundance patterns of the two components. As can be seen, there is an overall remarkable agreement with the literature data, with average differences at the 0.01-dex level. On an element-to-element basis, the differences are systematically below 0.05 dex. Considering the weighted (by the inverse variance) average of all the abundances with respect to hydrogen, [X/H], we obtain $\langle$$\Delta$[X/H]$\rangle$ (A--B) = +0.0209$\pm$0.0031 dex. This is compelling on statistical grounds (at a 6.7$\sigma$ significance level) and supports the slight enhancement in metals of the primary. This conclusion is strengthened by the fact that $\Delta$[X/H] (A--B) is systematically positive for the 20 elements (Fig.~\ref{fig_condensation_16_Cyg}). The stellar parameters of the two components are so close that, e.g., differential diffusion effects are not expected to modify this conclusion \citep{deal15}. The detection of such a small metallicity difference between the two components indicates that a precision at the 0.02-0.03 dex level is realistically achieved for the abundances of the solar analogues in our sample (once again, it does not pertain to stars outside of this category). It is an interesting conclusion in the context of large-scale spectroscopic surveys, as it shows that such subtle abundance differences can be revealed even with data of limited quality --- especially resolving power ---  provided that adequate analysis strategies are implemented. However, we are unable to find evidence for the increase of $\Delta$[X/H] (A--B) as a function of $T_\mathrm{c}$ reported by \citet[][]{tucci_maia19} and, to a lesser extent, by \citet{laws01} and N17. In contrast, the $\Delta$[X/H] (A--B)-$T_\mathrm{c}$ behaviour we find is completely flat (--0.21$\pm$0.79 $\times$ 10$^{-5}$ dex K$^{-1}$). This might be ascribed to our lower data quality \citep[but see, e.g.,][]{ramirez11}.

\end{appendix}

\end{document}